\documentclass[12pt]{article}
\usepackage{enumerate}
\usepackage{natbib}
\usepackage{url} 
\usepackage{xr}

\newcommand{\blind}{0}

\usepackage{psfrag,epsf}
\usepackage[utf8]{inputenc}
\usepackage{amsthm,amssymb,amsmath,amsfonts,amstext}
\usepackage[english]{babel}
\usepackage{array,latexsym}
\usepackage{graphicx,lscape}
\usepackage{enumitem}
\usepackage{multirow}
\usepackage[dvipsnames]{xcolor}

\newcommand{\var}{\operatorname{var}}

\usepackage{commath,mathrsfs,float,bbm,dsfont,soul,subfigure,tikz}
\usepackage{setspace,newlfont,authblk,lscape,longtable,multirow,comment,xfrac,indentfirst}
\usepackage[T1]{fontenc}
\usepackage[title]{appendix}
\newtheorem{prop}{Proposition}
\newtheorem{assump}{Assumption}
\newtheorem{theo}{Theorem}

\newcolumntype{G}{> {\columncolor[gray]{0.8}}c}

\usepackage{setspace} 

\begin{document}

\def\spacingset#1{\renewcommand{\baselinestretch}%
	{#1}\small\normalsize} \spacingset{1}


\if0\blind
{
	\title{\bf Estimating heterogeneous causal effects in observational studies using small area predictors}
	\author{Setareh Ranjbar\thanks{
			The authors gratefully acknowledge \textit{that the work of Nicola Salvati and Setareh Ranjbar has been partially carried out with the support of project InGRID 2 (grant agreement 730998, EU), and the work of Barbara Pacini and Nicola Salvati has been  partially carried out with the support of project PRA2018-9 (‘From survey-based to register-based statis- tics: a paradigm shift using latent variable models’). The work of Setareh Ranjbar has also benefited from the financial supports of the SNSF project (reference: 100018-178964). The authors are also grateful to Ray Chambers for his insight and constructive comments in writing the current version of the paper. }}\hspace{.2cm}\\
		Department of Business and Economics [HEC], University of Lausanne\\
		and \\
		Nicola Salvati \\
		Department of Economics and Management, University of Pisa\\
		and \\
		Barbara Pacini\\
		Department of Political Science, University of Pisa}
	\maketitle
} \fi

\if1\blind
{
	\bigskip
	\bigskip
	\bigskip
	\begin{center}
		{\LARGE\bf Estimating heterogeneous causal effects in observational studies using small area predictors}
	\end{center}
	\medskip
} \fi

\bigskip
\begin{abstract}
	The official statistics produced by National Statistical Institutes are mainly used by policy makers to take decisions. In particular, when policy makers and decision takers would like to know the impact of a given policy, it is important to acknowledge the heterogeneity of the treatment effects for different domains . If the domain of interest is small with regards to its sample size , then the evaluator has entered the small area estimation (SAE) dilemma. 
	Based on the modification of the Inverse Propensity Weighting estimator and the traditional small area predictors, we propose a new methodology to estimate area specific average treatment effects for unplanned domains.  A robustified version of the predictor against the presence of the outliers is also proposed. We develop analytical Mean Squared Error (MSE) estimators of the proposed predictors. By means of these methods, we can provide a map of policy impacts that can help to better target the treatment group(s). We illustrate the properties of these small area estimators by means of  a design-based simulation based on real data set where the target is to study the effects of permanent versus temporary contracts on the economic insecurity of households in different regions of Italy.
\end{abstract}

\noindent%
{\it Keywords:}   M-quantile regression, linear mixed models, potential outcome, inverse propensity scores, heterogeneity of effects.

\spacingset{1.45} 
\section{Introduction}\label{sec:Intro}
In recent years, the thrust of planning process has shifted from the macro to the micro level. There is a demand from administrators and policy planners for reliable estimates of various parameters at the micro level \citep{chandra2011}. In particular, policy makers and decision takers would like to know the impact of a given policy in certain unplanned geographic, socio-demographic, or socio-economic domains. Thus, they are faced with the problem of estimating heterogeneous causal effects. Unfortunately, very often it is not possible to design a randomized experiment, and observational data (from censuses, administrative archives and surveys that are not designed for the purpose) are used to evaluate the effects of the intervention. In some cases, large databases with baseline covariates are available and the assignment to treatment (the benefit received) is known, but there is not enough information on the outcome variables to be representative of the unplanned domains. In particular, direct estimates are not accurate because sample surveys are usually designed so that direct estimators for larger domains (states, regions - macro level) lead to reliable estimates. If the domain of interested for impact evaluation is small with regards to its sample size (or even zero in some domains), then the evaluator has entered the small area estimation (SAE) dilemma. Small area techniques provide official statistics using the survey samples and other sources of available information from which the estimators can borrow strength.

It is still surprising that no link has been established between the SAE literature and causal analysis that would allow for evaluating the impact of such a policy or decisions at a finer  population level. There are exceptions but with different intentions. \cite{chan2018} attempts to combine the strength of the two fields, causal inference and small area estimation, to provide more precise generalization of the randomized trials to the entire population. This paper uses model-based techniques borrowed from the SAE literature to get a better estimate of the average treatment effect in the sub-classification stratas, which are defined by the propensity scores, that have a sparse sample from the randomized experiment. There has been some statistical research on how to assess the generalizability of randomized trials to the target population in which it may be implemented (external validity). \cite{stuart:2001} propose the use of propensity-score-based metrics to quantify the similarity of the participants in a randomized trial and a target population. \cite{stuart:2015} provide a case study using one particular method, which weights the subjects in a randomized trial to match the population on a set of observed characteristics.  
Methods for assessing and enhancing external validity are just beginning to be developed.  These studies and SAE methods share the aim to generalise the sample treatment effect to the population. However, the heterogeneity of the effects in different sub-populations is usually out of the scope of external validity analysis. 

In this paper, we propose new methods to estimate the area specific average
treatment effects for small areas in observational studies. The main motivation behind
this is that such methods allow for local rather than universal policy advices. Another
advantage of our proposed method over existing ones is that in case there are no treated
units within the sample the classical approach will provide no estimate of the effect whereas
the small area techniques can be used to predict the effects even if the sample size of the
treated or controlled group is zero in the area of interest. We adopt the nested error
unit-level regression models \citep{battese:1988} and the M-quantile models \citep{chambers2006} to estimate propensity scores and the unobserved outcome for the
population. Then to estimate the area specific average treatment effects for unplanned
domains we propose a modification of the Inverse Probability Weighting estimator based
on the estimated propensity scores and predicted outcomes \citep{Rosenbaum1983,Hahn1998} and we prove that it is double robust.

Borrowing from some recent papers we report two examples of impact assessment based on observational data in which our methodological proposal could improve the accuracy of the results at a finer level (territorial classification or population subgroups).

Bachtrogler et al. (2020) analyze the impact of the European Union's Cohesion Policy (CP) on manufacturing firm growth. They aim to assess whether and to which extent the effects of the regional CP investments on supported manufacturing firms' performance vary across different territorial settings (European countries and NUTS-2 regions). The paper combines firm-level data with a set of territorial characteristics of NUTS-2 regions (data assembled from three databases).  Firms for which no NUTS-2 information is available were dropped out from the study and the sample was further reduced due to poor availability of outcome variables (change in value added, employment growth, and growth in productivity) for some firms.
		
Starting from a growing interest in studying the effectiveness of therapies in real-world conditions, Wendling et al. (2018)  compare methods based on observational data from health care databases (e.g., commercial claims data, electronic health records, and national registries) to estimate treatment effects that are supposed to be heterogeneous, e.g. different in subpopulations that are excluded or underrepresented in Random Control Trials (RCTs).

Depending on the outcome variables of interest (which could also be rare), it may be difficult to have follow-up data for a representative sample of the subgroups of interest. In such a case, small area estimation techniques can help reconstruct the outcome variable where it is missing, using covariates available in the health care databases.

To show the potential of our proposal, in this paper we consider a design-based simulation generating artificial data based on real data in an observational setting. Our experiment aims to approximate a real application in economic policy evaluation as closely as possible.

The paper is organised as follows. Section \ref{sec:Notation} is devoted to set out the theoretical background and the assumptions of the causal inference which is then used to extend the small area predictors. We introduce the proposed extensions to the Empirical Best Linear Unbiased Predictor (EBLUP) and M-quantile-based predictors under causal inference in Section \ref{sec:Estimator}. Their corresponding MSE estimators are presented in Section \ref{sec:MSE}. The performances of these newly proposed predictors are empirically assessed in Section \ref{sec:Application} by a design-based simulation based on EU-SILC data. Finally, in Section \ref{sec:conclusion} we summarise our main findings, and provide directions for future research. 



\section{Notation and assumptions}\label{sec:Notation}

To establish a common framework, we adopt capital letters for the outcome variable to take into account the probabilistic assignment mechanism to treatment,
unlike the conventional SAE notation where small letters are used to characterize a finite
population analysis. In what follows we use the bold cases to indicate vectors and matrices. The parameters of interest are shown using Greek letters , for example $\alpha$, and their estimates are distinct by carrying a `hat', for example $\hat{\alpha}$.

Consider a (super) population $\mathcal{U}$ of size $N$ that is partitioned into $m$ mutually disjoint sub-populations/domains $\mathcal{U}_{j}$ of size $N_{j}$, $j=1,\cdots, m$. In what follows we assume the availability of
survey data on the outcome variable and explanatory variables, which can be used to model it. In addition, the methods assume the availability of micro-level census or administrative data on the same set of explanatory variables. Therefore, we assume that values of a (continuous) outcome variable of interest $Y_{ij}$ are available from a random sample $s$, which includes units from all target domains. We assume that a set of auxiliary information, denoted as a vector of covariates $\mathbf{x}_{ij}$, is available for all the units in the population and that provides predictive power for the unobserved part of the population. It is also assumed that the vector $\mathbf{x}_{ij}$ of dimension $p \times 1$ contains the set of all confounders and some additional covariates that are useful in predicting the outcome. More generally, the vector of covariates may include both individual and area-level covariates.

 We are interested in studying the impact of a binary treatment, $W_{ij}$, that takes the value $ 1 $ for treated and $ 0 $ for non-treated (control) units in the population. We focus on treatment assigned at the individual level and assume that the information on treatment status exists for all population units, for example from administrative sources. This is a plausible assumption in many applications, such as unemployment benefits, government subsidies, pensions. 

We denote the sample size, the sampled part of the population and the non-sampled part of the population in each small area $j$ by $n_j$, $s_j$ and $r_j$ respectively, with $\mathcal{U}_j =s_j\bigcup r_j$. The total sample size is given by  $n=\sum_{j=1}^{m}n_{j}$.

To link the two methodologies on small area estimation and causal inference, we adopt the framework of Rubin Causal Model (RCM) \citep{Rubin74}, and use the approach of potential outcomes to properly define the causal estimands of interest. In small area estimation setting the aim is to provide estimates of the average effects for each small sub-population or domain (i.e., these are the unplanned domains in the survey) rather than for the entire population. This is particularly relevant when heterogeneous effects are expected among different domains. In these cases our proposal can provide a map of policy impacts at a small area level, helping to better understand the outcome of an intervention and to better target the treatment group(s).

The potential outcome approach is firstly developed under SUTVA \citep[Stable Unit Value Assumption;][] {Rubin1980}, stating that the outcome of each unit is unaffected by the treatment assignment of any other unit and also that there are no different versions of each treatment level, which may lead to different potential outcomes.  Within the simplest framework, each unit has only two potential outcomes, defined as $Y^{0}_{ij}$ and $Y^{1}_{ij}$ under control and under treatment, respectively. The former, $Y^{0}_{ij}$, denotes the outcome that would be realized by the individual if he or she is not treated and the latter, $Y^{1}_{ij}$, indicates the outcome that would be realized by the same individual if he or she is treated.  The potential outcomes for each unit would be vector-valued instead of scalars, including all the possible combinations of treatment assignment for a set of units.  

For the sampled units of area $j$ (the set $s_{j}$) only one of the potential outcomes is observed for each individual; the other is necessarily missing and needs to be predicted, entering the so called fundamental problem of causal inference.  We then observe the outcome variable $Y_{ij}$ where $Y_{ij}=W_{ij}Y^{1}_{ij}+(1-W_{ij}) Y^{0}_{ij}$, in this set.
For the non-sampled units of area $j$ (the set $r_{j}$), however, neither of the potential outcomes are  available and both are need to be predicted, implying that for the out of sample units $Y_{ij}$s are never observed. In this respect, our problem resembles that studied widely in the literature of imputation for missing data in the context of small area estimation. See \citet{Haziza2010}, \citet{Cantoni2018} and \citet{Chen2019} for a comprehensive review of this topic. The main difference of this line of literature with our work is twofold: (i) causal inference require additional assumptions and (ii) the percentage of missing values for which we need to predict the value is not negligible.  


The individual treatment effect for the unit $i$ in area $j$ can be defined as a comparison of potential outcomes, such as the difference, denoted as:
$$ \tau_{ij}=  Y^{1}_{ij}- Y^{0}_{ij}.$$
This parameter is not identifiable due to a lack of information for each unit, but several causal estimands can be defined as summaries of individual effects, which are identifiable and can be estimated out of the data under some additional assumptions.
Here we distinguish between two sets of estimands that are essential for our analysis. The first includes the average treatment effect (SATE)
for the sample units. The second set of estimands includes  the ATE 
for the population, named PATE.
Each of these estimands can be defined at the area (domain) level as follows:

\begin{equation} \label{eq:SATE}
\tau_{SATE_{j}}=\frac{1}{n_{j}}\sum_{i=1}^{n_{j}} \left(Y^{1}_{ij}- Y^{0}_{ij}\right)  ,
\end{equation}


\begin{equation} \label{eq:PATE}
\tau_{PATE_{j}}=\frac{1}{N_{j}}\sum_{i=1}^{N_{j}} \left(Y^{1}_{ij}- Y^{0}_{ij}\right).
\end{equation}


The aim of our proposal is to provide reliable estimates of $\tau_{PATE_{j}}$ 
for different areas/domains, borrowing strength from small area estimation  techniques. 

Causal effects from observational data can be identified under a set of assumptions, guaranteeing  that the treatment is effectively randomized within cells defined by the values of a set of observed covariates. Slight modifications are needed in some cases for the identification of heterogeneous effects among different domains. 

Here, we assume SUTVA, which is implied in the notation above, together with strong ignorability assumptions:   

\begin{assump} \label{as1}
Stable Unit Treatment Value
\end{assump}
The potential outcome for any unit does not vary with the treatments assigned to other units, and, for each unit, there are no different forms or versions of each treatment level. 
With multilevel data, this assumption may be questionable, especially for the units in the same area/domain. The implications of cluster structure, which may affect both the assignment to treatment and potential outcomes, have not been intensively studied, with a few exceptions \citep{Arpino2011, Li2013, Kim2017,Cafri2019}. However in our study, it is reasonable to assume that the treatment administered at the unit level will not affect other units  within the same area and that there are no expected movements and interference across domains.  Therefore, SUTVA will be maintained assuming no interference within and between clusters.

\begin{assump}\label{as2}
Unconfoundedness based on propensity scores
\end{assump}
 The assignment mechanism is unconfounded \citep[with the potential outcomes,][]{Rosenbaum1983} if:
$$ W_{ij} \perp(Y_{ij}^{1}, Y_{ij}^{0}) \mid \mathbf{X}_{ij}=\mathbf{x}_{ij}, \qquad \forall i \in \mathcal{U}_{j},$$
or
$$ W_{ij} \perp(Y_{ij}^{1}, Y_{ij}^{0}) \mid e(\mathbf{x}_{ij}), \qquad \forall i \in \mathcal{U}_{j},$$

where $e(\mathbf{x}_{ij}) = Pr(W_{ij}=1| \mathbf{X}_{ij}=\mathbf{x}_{ij})$ is known as a propensity score.
We assume that, conditional on a set of pre-treatment covariates or conditional solely on the propensity scores, the assignment mechanism is independent from the potential outcomes.


\begin{assump}\label{as3}
Common support (overlap)
\end{assump}
We assume that the unconfounded assignment mechanism is probabilistic, that is all the unit-level probabilities for receiving treatment are strictly between zero and one:
$$0 < e(\mathbf{x}_{ij}) = Pr(W_{ij}=1| \mathbf{X}_{ij}=\mathbf{x}_{ij}) < 1 \qquad \forall i \in \mathcal{U}_{j}.$$

In other words, each unit in the defined population has a chance of being treated and a chance of not being treated  \citep{Rosenbaum1983}. 
We assume common support within area, based on the whole set of population auxiliary variables. 

Treatment assignment mechanisms satisfying both overlap and unconfoundedness are called strongly ignorable, so that we assume strong ignorability within each area/domain.

\citet{Zanutto2004} discuss the importance of using propensity scores to match the treatment and control units while using regression models in the complex survey  settings. This approach can be also considered as a diagnostic tool to test the Assumption \ref{as3}.


\section{Small area estimators for causal inference}\label{sec:Estimator}
We propose a modification of the Augmented Inverse Probability Weighting estimator \citep{Rosenbaum1983}, based on the estimated propensity scores and predicted outcomes. As stated in the introduction, our main objective is to identify the heterogeneity of causal effects
among different areas in observational studies.

We start from defining the Average Treatment Effect in area $j$ as $\tau_{j}= E_{j}\left[\tau_{ij}\right]= E_{j} [Y_{ij}^{1}]- E_{j} [Y_{ij}^{0}]$  
, where the expectation is evaluated over the units in area $j$.
Under unconfoundedness Assumption 2, \cite{Imbens2009} show that:
$$ E_{j} [Y_{ij}^{1}] = E_{j}\left[\frac{W_{ij} Y_{ij}}{e(\mathbf{x}_{ij})}\right],$$
and
$$  E_{j} [Y_{ij}^{0}] = E_{j}\left[\frac{(1-W_{ij}) Y_{ij}}{1-e(\mathbf{x}_{ij})}\right],$$
where $e(\cdot)$ is the function that determines the units propensity scores (i.e., the prob-
ability for each unit to be treated) based on their vector of confounding covariates, $\mathbf{x}_{ij}$. The natural sample estimator for this parameter is:

\begin{equation*} \label{eq:SATE2}
\tilde{\tau}_{SATE_{j}}=\frac{1}{n_{j}} \sum_{i=1}^{n_{j}}\left[\frac{w_{ij}y_{ij}}{e(\mathbf{x}_{ij})}-\frac{(1-w_{ij})y_{ij}}{1-e(\mathbf{x}_{ij})}\right].
\end{equation*}
 If the propensity scores are unknown, the $e(\mathbf{x}_{ij})$ values need to be replaced by their estimates:
  \begin{equation*} \label{eq:SATE2Al}
 \tau_{SATE_{j}}^{\star}=\frac{1}{n_{j}} \sum_{i=1}^{n_{j}}\left[\frac{w_{ij}y_{ij}}{\hat{e}(\mathbf{x}_{ij})}-\frac{(1-w_{ij})y_{ij}}{1-\hat{e}(\mathbf{x}_{ij})}\right].
 \end{equation*}
\cite{Lunceford2004} and \cite{Imbens2004} propose to improve the performance of estimator \eqref{eq:SATE2Al} by re-normalizing the weights so that they some up to one:
 \begin{equation}\label{eq:SATE3}
 \hat{\tau}_{SATE_{j}} = \left(\sum_{i\in s_{j}}\left[\frac{w_{ij}y_{ij}}{\hat{e}(\mathbf{x}_{ij})}\right]\right)
 \left( \sum_{i=1}^{n_{j}}\frac{w_{ij}}{\hat{e}(\mathbf{x}_{ij})}\right)^{-1} - \left(\sum_{i\in s_{j}}\left[\frac{(1-w_{ij})y_{ij}}{1-\hat{e}(\mathbf{x}_{ij})}\right]\right) \left(\sum_{i=1}^{n_{j}}\frac{1-w_{ij}}{1-\hat{e}(\mathbf{x}_{ij})}\right)^{-1}.  
 \end{equation}

In the context of small area estimation we assume the unit level auxiliary information and the treatment status are available to predict the outcome and the propensity scores for the non-sampled part of the population in each small area. Therefore, we extend equation (\ref{eq:SATE2}) to the entire population providing an estimate of $\tau_{PATE_{j}}$ \eqref{eq:PATE} as it follows:

\begin{align} \label{eq:PATE3}
\hat{\tau}_{PATE_{j}} & =\left. \left(\sum_{i\in s_{j}}\left[\frac{w_{ij}y_{ij}}{\hat{e}(\mathbf{x}_{ij})}\right]+\sum_{i\in r_{j}}\left[\frac{w_{ij}\hat{y}_{ij}}{\hat{e}(\mathbf{x}_{ij})}\right]\right) \left(\sum_{i=1}^{N_{j}}\frac{w_{ij}}{\hat{e}(\mathbf{x}_{ij})} \right)^{-1} \right. - \nonumber \\
&
\left. \left(\sum_{i\in s_{j}}\left[\frac{(1-w_{ij})y_{ij}}{1-\hat{e}(\mathbf{x}_{ij})}\right]+\sum_{i\in r_{j}}\left[\frac{(1-w_{ij})\hat{y}_{ij}}{1-\hat{e}(\mathbf{x}_{ij})}\right]\right)\left(\sum_{i=1}^{N_{j}}\frac{1-w_{ij}}{1-\hat{e}(\mathbf{x}_{ij})}\right)^{-1}\right. .
\end{align}

In Theorem \ref{theo:doub-rob} we show that the estimator \eqref{eq:PATE3} has consistent and double robust properties without any extra adjustment. On the contrary, if in equation \eqref{eq:PATE3} the weights are not re-normalized the estimator is no more double robust and an adjustment term is needed to obtain this desirable property, see the Section \ref{S-sec:doub_robust} in supplementary materials.

In what follows we refer to (\ref{eq:SATE3}) as the IPW-Direct estimator, which is the classical Inverse Propensity Weighting estimator proposed by \cite{Rosenbaum1983}. Alternative direct estimators, that use the survey weights, have been proposed by \cite{Zanutto2006} and by \citet{Miratrix2018}. 

We can show the consistency and double robust properties of the proposed estimator in equation \eqref{eq:PATE3} by expressing it as a weighted average of the outcomes $y_{ij}$s and their estimates $\hat{y}_{ij}$s in each small area:

\begin{equation*} \label{eq:PATE_final}
\hat{\tau}_{PATE_{j}} =\left(\sum_{i\in s_{j}}a_{ij}y_{ij}+\sum_{i\in r_{j}}a_{ij}\hat{y}_{ij}\right) \left(\sum_{i=1}^{N_{j}} a_{ij}\right)^{-1} - \left(\sum_{i\in s_{j}}b_{ij}y_{ij}+\sum_{i\in r_{j}}b_{ij}\hat{y}_{ij}\right) \left(\sum_{i=1}^{N_{j}} b_{ij}\right)^{-1},
\end{equation*}
where $\{a_{ij}=\frac{w_{ij}}{\hat{e}(x_{ij})}\}$ and $\{b_{ij}=\frac{1- w_{ij}}{1-\hat{e}(x_{ij})}\}$ are the sequences of weights of area $j$. This allows to prove that $\hat{\tau}_{PATE_{j}}$, as a weighted average of i.i.d random variables (rvs) conditioned on the small area $j$, is double robust  and consistent as $N_j \rightarrow \infty$.
Let $A_{N_{j}}$ and $B_{N_{j}}$ be $\sum_{i \in U_{j}}a_{ij}$ and $\sum_{i \in U_{j}}b_{ij}$, respectively. The theory will be developed under Assumptions 1, 2, 3. Further the following conditions have to be satisfied for the convergence of $\hat{\tau}_{PATE_{j}}$ to its true value:
\begin{itemize}
	\item[(a)] $A_{N_{j}} \rightarrow \infty $ and $a_{ij}/A_{N_{j}} \rightarrow 0;$ 
	\item[(b)] $B_{N_{j}}\rightarrow \infty $ and $b_{ij}/B_{N_{j}} \rightarrow 0;$
	\item[(c)] $a_{ij}$s and $b_{ij}$s are bounded;
	\item[(d)] the $\var(Y_{ij}) < \infty$ and $\var(Y_{ij}-\hat{Y}_{ij}) < \infty$.
\end{itemize}
Assumption \ref{as2} and the fact that the propensity scores are estimated using the information on the whole population guarantee that weights are a deterministic sequence of values given the area population of size $N_j$. The conditions that  $A_{N_{j}} \rightarrow \infty $ and $B_{N_{j}}\rightarrow \infty $ are linked with the assumptions that in each area there must be treated and non-treated units in the population. While it is possible to provide the estimates even if the entire sample units in an area belong only to treated or control group, at the population level the presence of both groups is essential to provide area level estimates of the treatment. Further, according to Assumption \ref{as3} the propensity scores take values between 0 and 1 away from the boundaries. When $N_j \rightarrow \infty$, where the propensity scores are bounded away from 0, $a_{ij}$s  are bounded and $a_{ij}/A_{N_{j}} \rightarrow 0$. Likewise, when the scores are bounded away from 1, $b_{ij}$s are bounded and $b_{ij}/B_{N_{j}} \rightarrow 0$. 
\begin{theo} \label{theo:doub-rob}
	Under assumptions \ref{as1}-\ref{as3} and the conditions (a), (b), (c) and (d) the estimator \eqref{eq:PATE3} is double robust and consistent. That is:
$$ Pr\left(\lim\limits_{N_{j}\rightarrow \infty}\hat{\tau}_{PATE_{j}} =	E_j\left[Y^{1}_{ij}\right]- E_j\left[Y^{0}_{ij}\right]\right)=1,$$  
	\begin{itemize}
		\item[(i)] as long as the propensity score model is correct, even if the postulated prediction model is incorrect;
		\item[(ii)]as long as the prediction model is correct, even if the postulated propensity model is incorrect.	 
	\end{itemize}
\end{theo}

The proof of Theorem 1 is in Section \ref{S-sec_proofth}, supplementary materials. In estimator (\ref{eq:PATE3}) different methods can be adopted to predict the unobserved $y_{ij}$s and to estimate the propensity scores.Here we propose two approaches for estimation strategies and discuss the resulting impact on the estimation of \eqref{eq:PATE3}. In the first proposal we predict the unobserved outcomes using EBLUP and a generalized linear mixed model to estimate the propensities. This estimator is referred to as  IPW-EBLUP hereafter and can also be seen as a modification of the EBLUP estimator for the area level mean.  In the second proposal we use a robust approach based on M-quantile models proposed by \cite{chambers2006} for the continuous outcome and by \cite{Chambers2016} for the binary case to predict the unobserved outcomes and estimate the propensity scores. The resulting estimator is labelled IPW-MQ hereafter. We explain in more detail the models and the estimating strategies used for IPW-EBLUP and IPW-MQ in Section \ref{sec:estimation}.

The properties of IPW-Direct estimators are widely studied in the literature; see for instance \cite{Hirano2003} and \cite{Wooldridge2007} for more details. However,
when the area/domain sample sizes are small these estimates are no longer reliable at this fine levels, that is, they could vary significantly. Our proposed estimators IPW-EBLUP and IPW-MQ overcome this problem by borrowing strength from additional sources of information rather than merely using the sample data. The second estimator can also deal with data that is contaminated by outlying values.  

\subsection{Data generating processes and estimation strategies}\label{sec:estimation}

To explain the data generating process and justify our estimation strategies for predicting the unobserved population outcomes and estimating the population propensity scores once again we use the potential outcome framework. Consider the two potential outcomes for individual $i$ in area $j$ to be related in the following way \cite[p.~263]{ImbRub15}:
$$ Y_{ij}^{1}=Y_{ij}^{0}+\tau_{j}, $$
where $\tau_{j}$ is the area specific causal effect of a policy intervention. Since our main objective is to acknowledge the heterogeneity of the average treatment effect over  sub-populations (here small areas) we do not take into account the individual level heterogeneity, i.e.\ $\tau_{PATE_{j}}=\tau_{SATE_{j}}=\tau_{j}. $

To benefit from the hierarchical structure in the data, {\color{black} without loss of generality}, we consider {\color{black} a nested error} linear model \citep{battese:1988} as the data generating process of the potential outcome in the absence of the treatment:
$$ y_{ij}^{0}=\mathbf{x}_{ij}^{T}\boldsymbol{\beta}+u_{j}+\epsilon_{ij},$$
where $u_{j}$ is the area specific random effect and $\epsilon_{ij}$ is the individual errors, the distributions of which are to be assumed (in general normal) if the model is fitted parametrically. This holds for the entire population as well as for the sample at hand in the absence of sample selection bias. Let $w_{ij}$ be the individual treatment status, the outcome (observed in the sample and not observed for the population) is:
\begin{align}\label{model_mixed}
\nonumber y_{ij} & = (w_{ij})y_{ij}^{1}+(1-w_{ij})y_{ij}^{0} \\
& = \mathbf{x}_{ij}^{T}\boldsymbol{\beta}+w_{ij}\tau_{j}+u_{j}+\epsilon_{ij}.  
\end{align}
In the context of small area estimation we need to fit this model to the sample data and predict the outcome for the entire population by using the estimated parameters of the model and the auxiliary information that is available for the entire population. There are many different techniques that are developed in the SAE literature; two sets of parametric models are discussed in this paper, but, of course, others can also be adopted if appropriate. It is also worth noting that, if we have the area level variables in the model then the interaction between these variables and the treatment variable must also be included in the random part of the model \citep{Arpino2011}.  

\subsubsection{Out-of-sample estimation of outcome and propensities in hierarchical structure} \label{subsec:EBLUP-est}
We start by assuming that the area specific causal effects, $\tau_{j}$s, are randomly distributed with $\tau_{j} \sim \mathcal{N}(\gamma_{0},\sigma_{\gamma}^{2})$.  Then equation \eqref{model_mixed} can be rewritten as
\begin{equation}\label{eq:est2}
y_{ij}=\mathbf{\tilde{x}}_{ij}^{T}\boldsymbol{\tilde{\beta}}+w_{ij}\gamma_{j}+u_{j}+\epsilon_{ij},
\end{equation}
where $\mathbf{\tilde{x}}_{ij}=(\mathbf{x}_{ij}^{T},w_{ij})^{T}$ is of dimension $(p+1)\times 1$, $ \boldsymbol{\tilde{\beta}}=(\boldsymbol{\beta}^{T},\gamma_{0})^{T}$ is the vector of fixed effects and we further assume that $ u_{j}\sim \mathcal{N}(0,\sigma_{u}^2)$, and $\epsilon_{ij}\sim \mathcal{N}(0,\sigma_{\epsilon})$. As a consequence of our assumption on the distribution of the area specific causal effects we have $\gamma_{j}\sim \mathcal{N}(0,\sigma_{\gamma})$, that is the random slope associated with the treatment status.
For obtaining the IPW-EBLUP, a mixed linear model (more specifically a random slope model) is fitted, using the maximum likelihood (ML) or restricted maximum likelihood (REML) method  \citep{McCulloch2001,Pinheiro2006}. Then the estimated parameters are used to predict the outcome $\hat{y}_{ij}$ for $i \in r_{j}$ under model \eqref{eq:est2}. The assumption of normality of the random components are mainly in place to specify the form of ML or REML used for estimating the unknown parameters of the model, including the unknown parameters of the variance-covariance matrix. However, this assumption can easily be relaxed using other existing methods for fitting random effect models, such as quasi-likelihood methods or Generalized Estimating Equation \citep{Liang1986} under some other mild conditions.  
\begin{prop}
Under Assumption \ref{as2}, unconfoundedness, the vector of random slopes $\boldsymbol{\gamma}$ and random intercepts $\boldsymbol{u}$ in equation \eqref{eq:est2} are independent, that is:
$$\begin{bmatrix} \boldsymbol{\gamma} \\ \boldsymbol{u} \end{bmatrix} \overset{i.i.d}{\sim } \begin{pmatrix} \mathbf{0}, &  \boldsymbol{\Sigma}_{\omega}\end{pmatrix},$$
where $\boldsymbol{\gamma}=(\gamma_{1}, \cdots, \gamma_{m})^{T}$, $\boldsymbol{u}=(u_{1}, \cdots , u_{m})$, and $\boldsymbol{\Sigma}_{\omega}=\begin{pmatrix} \boldsymbol{\Sigma}_{\gamma} & \mathbf{0}\\
\mathbf{0} & \boldsymbol{\Sigma}_{u}\end{pmatrix}$.
\end{prop}
\begin{proof} Based on Assumption \ref{as2} the treatment assignment is independent from the potential outcomes conditional on the set of pre-treatment covariates (confounders). This assumption requires that conditional on observed covariates there are no unobserved factors that are associated both with the assignment mechanism and potential outcomes, that is,  $E\left[\gamma_{j}\left(u_{j}+\epsilon_{ij}\right)\right]=0$. Because $E\left[\gamma_{j}\epsilon_{ij}\right]=0$ it goes that $E\left[\gamma_{j}u_{j}\right]=0$.
\end{proof}


In equation \eqref{eq:est2} the average return to $w_{ij}$ is captured by the fixed effect and the area specific heterogeneity of the return  to $w_{ij}$ is modeled through a random slope $\gamma_{j}$ , that needs to be predicted \citep{Li2013}. However, our estimators of the total causal effect do not merely depend on the estimation/prediction of these two effects. In addition, we balance the characteristics of treated and control groups by weighting the outcomes based on the individual propensity scores. Therefore, these estimators have doubly robust properties \citep{Bang2005}, that is, having misspecified only one of the models for the prediction of the outcomes or for the estimation of the propensity scores, we can still provide a consistent estimator for the causal effects of each area. Further, the hierarchical structure of the data as it is defined in equation \eqref{eq:est2} for the outcome model should also be considered in the estimation model of the propensity scores, see \citet{Arpino2011} and \citet{Arpino2016}. Then, we consider the following model for the propensity scores:
\begin{equation}\label{eq:est3}
\eta_{ij}=\Lambda(e(\mathbf{x}_{ij}))= \mathbf{x}_{ij}^{T}\boldsymbol{\alpha}+\nu_{j},
\end{equation}
where $\Lambda(.)$ is a logit link function. Substituting the estimated values $\hat{y}_{ij}=\mathbf{\tilde{x}}_{ij}^{T}\hat{\tilde{\boldsymbol{\beta}}}+w_{ij}\hat{\gamma}_{j}+\hat{u}_{j}$ and $\hat{e}(x_{ij})=\Lambda^{-1}(\mathbf{x}_{ij}^{T}\hat{\boldsymbol{\alpha}}+\hat{\nu}_{j})$ in equation \eqref{eq:PATE3} provides the estimates of IPW-EBLUP. 

\subsubsection{Robust estimation for out-of-sample units}\label{subsec: MQ_est}
An alternative to mixed models and IPW-EBLUP is given by the M-quantile regression models for estimating the outcome variable and the propensity scores. If an outlying value can destabilize a population estimate based on a large
survey sample, it can almost certainly destroy the validity of the corresponding direct estimate for the small area from which the outlier is sourced, since this estimate will be based on a much smaller sample size. This problem does not disappear when the small area estimator is a model based estimator such as EBLUP: large deviations from the expected response (outliers) are known to have a large influence on classical maximum likelihood inference based on generalized linear mixed models (GLMM).  \citet{chambers2006} and \citet{sinha:2009} addressed the issue of outlier robustness in SAE proposing techniques that can be used to down-weigh any outliers when fitting the underlying model. In particular, \citet{chambers2006} proposed to apply the M-quantile regression models to SAE with the aim of obtaining reliable and outlier robust estimators without recourse to parametric assumptions for the residuals distribution using M-estimation theory. For details on M-quantile regression see \cite{Breckling:1988}.

When using the M-quantile method the unobserved outcomes are predicted as follows: 
\begin{equation}\label{eq:estMQ1}
\hat{y}_{ij}= \mathbf{x}_{ij}^{T}\hat{\boldsymbol{\beta}}_{\hat{\bar{q}}_{j}}+ w_{ij} \hat{\gamma}_{\hat{\bar{q}}_{j}}, 
\end{equation}

where $\boldsymbol{\hat{\beta}}_{\bar{q}_{j}}$ and $\hat{\gamma}_{\bar{q}_{j}}$ are the regression coefficients of the M-quantile  model estimate at quantile $\hat{\bar{q}}_{j}$ , that is, the average of the estimated quantiles for the sample units in area $j$. The \citet{chambers2006} proposal is an alternative to the random effect models for characterizing the variability across the population not accounted for by the regressors based on the the M-quantile coefficients {\color{black} of the population units}. The authors observed that if a hierarchical structure does explain part of the variability in the population data, units within areas defined by this hierarchy are expected to have similar M-quantile coefficients. For details on the computation of M-quantile coefficients see \cite{chambers2006}.

For estimating the propensity scores the M-quantile for binary data proposed by \cite{Chambers2016} is adopted. Modelling the M-quantiles of a binary outcome presents more challenges than modelling the M-quantiles of a count outcome. A detailed account of these challenges is provided in \cite{Chambers2016}. The authors proposed a new semiparametric M-quantile approach to small area
prediction for binary data that extends the ideas of \cite{Cantoni:2001} and \cite{chambers2006}. This predictor can be viewed as an outlier robust alternative to the more commonly used conditional expectation predictor \eqref{eq:est3} for binary data that is based on a logit GLMM with Gaussian random effects. With the proposed approach random effects are avoided and between-area variation in the response is characterized by variation in area-specific values of M-quantile indices. Furthermore, outlier robust inference is achieved in the presence of both misclassification and measurement error.

Under the M-quantile framework the propensity scores are estimated as: 
\begin{equation}\label{eq:estMQ2}
\hat{\eta}_{ij}=\Lambda(\hat{e}(\mathbf{x}_{ij}))= \mathbf{x}_{ij}^{T}\boldsymbol{\hat{\alpha}}_{\hat{\bar{q}}_{j}},    
\end{equation}
where the area level M-quantile coefficients are computed in different way with respect to the continuous outcome. See \cite{Chambers2016} for details.
Substituting the $\hat{y}_{ij}$ and $ \hat{e}(x_{ij})$ in equation (\ref{eq:PATE3}) provides the estimates of IPW-MQ. Note that this estimator is a special case of the equation \eqref{eq:PATE3} and so it is double robustness and consistent.   

\section{MSE estimators in the finite population}\label{sec:MSE}
In the context of randomized experiments \citet{Ding2019} proposed the decomposition of overall treatment effect variation into systematic and idiosyncratic components.  In this paper we are in the framework of observational data and we are using the inverse propensity weighting; for this reason, we decompose the variation of the effect into the variation due to the estimation of the (i) outcome and the (ii) propensity scores. For the first component of variation we propose its
estimation with an analytical derivation. In particular, for the IPW-EBLUP the proposal is based on the MSE estimation approach that is described in \citet{Prasad1990} and represents an extension of the ideas in \citet{opsomer2008}. For IPW-MQ the MSE estimator is based on second order approximations to the variances of solutions of outlier robust estimating equations and represents an extension of the ideas in \citet{Chambers2014}. The proposed analytical MSE estimators do not take into account the variability due to the estimation of the propensity scores. So to add this component of variability we suggest using re-sampling techniques.  \citet{Miratrix2018} point out the importance of considering the extra variability that is introduced when estimating $\tau_{PATE}$ using weights, which is a similar problem to ours. In particular, 
for IPW-EBLUP we suggest using a parametric bootstrap technique, such as that proposed by \cite{Gonzalez-Manteiga:2008} or a non-parametric bootstrap procedure as in \cite{opsomer2008}. For IPW-MQ, we suggest applying an outlier robust bootstrap estimator that is the  modified version of the block-bootstrap approach of \cite{Chambers:2013}. These bootstrap methods are explained in detail in Supplementary Material, Section  \ref{S-sec:boots-both}.



We show in model-based simulation experiments (Section \ref{S-sec:supp_Simulation}) how these approaches can be useful for estimating the MSE of various small area predictors that are considered in this paper.

To develop the analytical MSE estimators for small area predictors based on EBLUP and MQ approaches, we rewrite the estimator in equation (\ref{eq:PATE3}) as a linear combination of observed and unobserved outcomes:  

\begin{align}\label{eq:IPW2}
\hat{\tau}_{PATE_{j}} & =K_{j}^{-1}\left. \left(\sum_{i\in s_{j}}\left[\frac{w_{ij}y_{ij}}{\hat{e}(\mathbf{x}_{ij})}\right]+\sum_{i\in r_{j}}\left[\frac{w_{ij}\hat{y}_{ij}}{\hat{e}(\mathbf{x}_{ij})}\right]\right)\right. \nonumber -\\
&
T_{j}^{-1}\left. \left(\sum_{i\in s_{j}}\left[\frac{(1-w_{ij})y_{ij}}{1-\hat{e}(\mathbf{x}_{ij})}\right]+\sum_{i\in r_{j}}\left[\frac{(1-w_{ij})\hat{y}_{ij}}{1-\hat{e}(\mathbf{x}_{ij})}\right]\right)\right.\nonumber \nonumber \\
& = \sum_{i\in s_{j}}D_{ij}y_{ij}+\sum_{i\in r_{j}}D_{ij}\hat{y}_{ij},
\end{align}

where
$K_{j}= \sum_{i=1}^{N_{j}}w_{ij}/\hat{e}(\mathbf{x}_{ij})$,   $T_{j}=\sum_{i=1}^{N_{j}}(1-w_{ij})/(1-\hat{e}(\mathbf{x}_{ij}))$, and 
$$ D_{ij}= \left(\frac{K_{j}^{-1}w_{ij}}{\hat{e}(\mathbf{x}_{ij})}-\frac{T_{j}^{-1}(1-w_{ij})}{1-\hat{e}(\mathbf{x}_{ij})}\right).$$

\subsection{MSE of the causal effect estimator IPW-EBLUP}
We start from equation (\ref{eq:IPW2}) to derive the analytic formula of the MSE for IPW-EBLUP. We consider that the $D_{ij}$s are known for the entire population, so we do not account for their variations originating from the estimation of the propensity scores. 
If the proportion of observed outcomes, $f_{j}=\frac{n_{j}}{N_{j}}$, is small (negligible) we can write:  
$$ \hat{\tau}_{PATE_{j}}-\tau_{j}= \mathbf{D}_{j}^{T}\mathbf{\hat{y}}_{j}-\mathbf{D}_{j}^{T}\mathbf{y}_{j}= \mathbf{D}_{j}^{T}\left(\mathbf{\hat{y}}_{j}-\mathbf{y}_{j}\right). $$

where $\mathbf{D}_{j}$, $\mathbf{\hat{y}}_{j}$ and $\mathbf{y}_{j}$ are the vectors of $D_{ij}$s, the response variable and predicted outcomes, respectively, for the population in area $j$ \citep{Prasad1990}. The prediction of the outcome is obtained using the equation \eqref{eq:est2}:
\begin{equation}\label{pred_area} \mathbf{\hat{y}}_{j}= \mathbf{\tilde{X}}_{j}^{T}\boldsymbol{\hat{\tilde{\beta}}}+\mathbf{\tilde{W}}_{j}\boldsymbol{\hat{\gamma}}+ \mathbf{Z}_{j}\mathbf{\hat{u}},\end{equation}
where $\mathbf{\tilde{X}}_{j}$ is the matrix of auxiliary variables for area $j$ of dimension $(p+1) \times n_j$, $\mathbf{\tilde{W}}_{j}$ is a sparse matrix with the $j$th column being replaced by the treatment status of individuals in area $j$, $\mathbf{Z}_{j}$ is a sparse matrix of area indicators with only the elements of column $j$th equal to one, so that $\var(\mathbf{y})=\mathbf{V}= \mathbf{\tilde{W}}\boldsymbol{\Sigma}_{\gamma}\mathbf{\tilde{W}}^{T}+\mathbf{Z}\boldsymbol{\Sigma}_{u}\mathbf{Z}^{T}+\boldsymbol{\Sigma}_{\epsilon}$. If the variances of the random components are known, standard results from BLUP theory \citep[][Chapter 9]{McCulloch2001} guarantee that, given the model specifications \eqref{eq:est2} and Preposition 1, the generalized least squares estimator
$$ \boldsymbol{\hat{\tilde{\beta}}}= \left(\mathbf{\tilde{X}}^{T}\mathbf{V}^{-1}\mathbf{\tilde{X}}\right)^{-1}\mathbf{\tilde{X}}^{T}\mathbf{V}^{-1}\mathbf{Y}$$
and the predictors
$$\boldsymbol{\hat{\gamma}}= \boldsymbol{\Sigma}_{\gamma}\mathbf{\tilde{W}}^{T}\mathbf{V}^{-1}\left(\mathbf{Y}-\mathbf{\tilde{X}}\boldsymbol{\hat{\tilde{\beta}}}\right) $$
$$\mathbf{\hat{u}}= \boldsymbol{\Sigma}_{u}\mathbf{Z}^{T}\mathbf{V}^{-1}\left(\mathbf{Y}-\mathbf{\tilde{X}}\boldsymbol{\hat{\tilde{\beta}}}\right)$$
are optimal among linear estimators and predictors, respectively. Replacing $\mathbf{\hat{y}}_{j}$ with \eqref{pred_area} we can write
\begin{align} 
\nonumber \hat{\tau}_{PATE_{j}}-\tau_{j} &=  \mathbf{D}_{j}^{T}\left(\mathbf{\hat{y}}_{j}-\mathbf{y}_{j}\right)\\
\label{diff}& = \mathbf{D}_{j}^{T} \mathbf{c}_{j}(\boldsymbol{\hat{\tilde{\beta}}}-\boldsymbol{\tilde{\beta}})+\mathbf{D}_{j}^{T}\mathbf{\tilde{Z}}_{j}\left[\boldsymbol{\Sigma}_{\omega}\mathbf{\tilde{Z}}^{T}\mathbf{V}^{-1}\left(\mathbf{Y}-\mathbf{\tilde{X}}\boldsymbol{\tilde{\beta}}\right)- \omega\right],
\end{align}
where $\mathbf{c}_{j}=\mathbf{\tilde{X}}_{j}^{T}-\left(\mathbf{\tilde{Z}}_{j}\boldsymbol{\Sigma}_{\omega}\mathbf{\tilde{Z}}^{T}\mathbf{V}^{-1}\mathbf{\tilde{X}}\right)$, $\mathbf{\tilde{Z}}_{j}=(\mathbf{\tilde{W}}_{j},\mathbf{Z}_{j})$, $\mathbf{\tilde{Z}}=(\mathbf{\tilde{W}},\mathbf{Z})$, and $\boldsymbol{\omega}=\left(\boldsymbol{\gamma}^{T},\mathbf{u}^{T}\right)^{T}$, $\boldsymbol{\Sigma}_{\gamma}=\sigma_{\gamma}\mathbf{I}_{m}$, $ \boldsymbol{\Sigma}_{u}=\sigma_{u}\mathbf{I}_{m}.$ 
If both the random slopes and the area specific intercepts are treated as true random effects in the underlying model \eqref{eq:est2}, the mean prediction error is $0$ and the covariance between the two terms in equation \eqref{diff} is also 0, so that the MSE of the prediction errors is
\begin{equation} \label{eq:MSE1} E\left[(\hat{\tau}_{PATE_{j}}^{BLUP}-\tau_{j})^2\right] =  \mathbf{D}_{j}^{T}\mathbf{\tilde{Z}}_{j}\boldsymbol{\Sigma}_{\omega}\left(\mathbf{I}-\mathbf{\tilde{Z}}^{T}\mathbf{V}^{-1}\mathbf{Z}\boldsymbol{\Sigma}_{\omega}\right)\mathbf{\tilde{Z}}_{j}^{T}\mathbf{D}_{j}+\mathbf{D}_{j}^{T} \mathbf{c}_{j}(\mathbf{\tilde{X}}^{T}\mathbf{V}^{-1}\mathbf{\tilde{X}})\mathbf{c}_j^{T}\mathbf{D}_{j}.
\end{equation}
To extend these results for IPW-EBLUP, that is, where $\mathbf{V}$ is unknown, the variation that comes from the estimation of variance components has to be added. The resulting EBLUP version of equation \eqref{diff} is
\begin{equation}
    \mathbf{D}_{j}^{T}\hat{\mathbf{c}}_{j}(\boldsymbol{\hat{\tilde{\beta}}}-\boldsymbol{\tilde{\beta}})+\mathbf{D}_{j}^{T}\mathbf{\tilde{Z}}_{j}\left[\widehat{\boldsymbol{\Sigma}}_{\omega}\mathbf{\tilde{Z}}^{T}\widehat{\mathbf{V}}^{-1}\left(\mathbf{Y}-\mathbf{\tilde{X}}\boldsymbol{\tilde{\beta}}\right)- \omega\right],
\end{equation}
with $\hat{\mathbf{c}}_{j}=\mathbf{\tilde{X}}_{j}^{T}-\left(\mathbf{\tilde{Z}}_{j}\widehat{\boldsymbol{\Sigma}}_{\omega}\mathbf{\tilde{Z}}^{T}\widehat{\mathbf{V}}^{-1}\mathbf{\tilde{X}}\right)$ using restricted maximum likelihood estimators for the unknown variance components in $\mathbf{V}$ and $\mathbf{\Sigma}_{\omega}$. To derive a second-order approximation for the MSE as well as an estimator for the MSE that is correct up to the second order we follow the method proposed by \citet{opsomer2008}. The vector of unknown components of the variance-covariance matrix is $\boldsymbol{\theta}=(\sigma^{2}_{\gamma},\sigma^{2}_{u},\sigma^{2}_{\epsilon})$ and we define
 $$ \mathcal{S}_{t}= \mathbf{D}_{j}^{T}\mathbf{\tilde{Z}}_{j}\left(\frac{\partial\boldsymbol{\Sigma}_{\omega}}{\partial(\boldsymbol{\theta})_{t}}\mathbf{\tilde{Z}}^{T}\mathbf{V}^{-1}+\boldsymbol{\Sigma}_{\omega}\mathbf{\tilde{Z}}^{T} \frac{\partial\mathbf{V}^{-1}}{\partial(\boldsymbol{\theta})_{t}}\right),\qquad t=1,2,3.$$
 Further, let us define the $3 \times 3$ matrix $\mathcal{I}$, the Fisher information matrix with respect to the variance components $\boldsymbol{\theta}$, then, the MSE of the IPW-EBLUP predictor is given by
 \begin{equation}\label{eq:MSE2}
    MSE(\hat{\tau}_{PATE_{j}}^{EBLUP})= E\left[(\hat{\tau}_{PATE_{j}}^{BLUP}-\tau_{j})^2\right]+tr\left(\mathcal{S} \mathbf{V}\mathcal{S}^{T} \mathcal{I}^{-1}\right) + o(m^{-1}),
 \end{equation}
 and its estimator can be obtained as
 \begin{align} \nonumber mse(\hat{\tau}_{PATE_{j}}^{EBLUP}) =& \mathbf{D}_{j}^{T}\mathbf{\tilde{Z}}_{j}\widehat{\boldsymbol{\Sigma}}_{\omega}\left(\mathbf{I}-\mathbf{\tilde{Z}}^{T}\widehat{\mathbf{V}}^{-1}\mathbf{\tilde{Z}}\widehat{\boldsymbol{\Sigma}}_{\omega}\right)\mathbf{\tilde{Z}}_{j}^{T}\mathbf{D}_{j}+\mathbf{D}_{j}^{T} \hat{\mathbf{c}}_j(\mathbf{\tilde{X}}^{T}\widehat{\mathbf{V}}^{-1}\mathbf{\tilde{X}})\hat{\mathbf{c}}_j^{T}\mathbf{D}_{j}\\ \label{eq:mse-eblup}
 & + 2\left(\mathbf{Y}-\mathbf{\tilde{X}}\boldsymbol{\hat{\tilde{\beta}}}\right)^{T} \mathcal{\widehat{S}}^{T} \hat{\mathcal{I}}^{-1}\mathcal{\widehat{S}}\left(\mathbf{Y}-\mathbf{\tilde{X}}\boldsymbol{\hat{\tilde{\beta}}}\right),
 \end{align}
 substituting $\boldsymbol{\theta}$ by the restricted maximum likelihood estimates in $\mathcal{S}$ and $\mathcal{I}$. Using some results of this section, asymptotic properties of the IPW-EBLUP are obtained.
 \begin{prop} \label{prop:eblup}
Under assumptions 1-3 and the conditions (a), (b), (c) and (d) of Theorem \ref{theo:doub-rob} and the normality assumption on the random effects and the error terms, the estimator IPW-EBLUP is double robust and asymptotically normally distributed:
	$$ \sqrt{N_{j}m}(\hat{\tau}_{PATE_j}^{EBLUP}-\tau_j )\sim \mathcal{N}(0, \mathscr{V}_{j}(\theta)).$$
	as $m\rightarrow \infty$.
	\end{prop}
The proof of Preposition \ref{prop:eblup} is provided in the Section \ref{S-sec:asym:eblup} of the Supplementary Material.
 
\subsection{MSE of the robust causal effect estimator IPW-MQ}
In this section we propose an analytical derivation of the MSE for the IPW-MQ type estimator. This is based on the linearization ideas that are set out in \cite{Booth:1998} and that are used by \cite{Chambers2014} to propose a new estimator of the MSE of a small area estimator that is defined by the
solution of a set of robust estimating equations. The MSE is a sum of a prediction variance and a squared bias term. The theoretical development, as in \cite{Chambers2014}, is based on approximations that correspond to assuming that $max(n_{j})=O(1)$, so that, as the number of small areas tends to infinity, the prediction variance and the squared bias are $O(1)$. We also make the standard assumption that a consistent estimator of the MSE of a linear approximation to the small area estimator of interest can be used as its MSE estimator. As noted by \cite{Harville:1992}, such an approach
will not generally be consistent, and the resulting MSE estimator can be downward biased. However, in small
sample problems, this is not generally an issue.

Note that we assume that the $\bar{q}_j$  values are known. The prediction error of the IPW-MQ estimator is then:
\begin{equation}\label{prediction_error}
\hat{\tau}_{PATE_j}^{MQ}-\tau_j=\sum_{i \in r_j}D_{ij}\hat{y}_{ij}-\sum_{i \in r_j}D_{ij}{y}_{ij},
\end{equation}
where $\hat{y}_{ij}= \mathbf{x}_{ij}^{T}\hat{\boldsymbol{\beta}}_{\bar{q}_{j}}+ w_{ij} \hat{\gamma}_{\bar{q}_{j}}$. Following \cite{Chambers2014} the prediction variance of IPW-MQ estimator is:
\begin{equation}\label{prediction_var}
\var(\hat{\tau}_{PATE_j}^{MQ}-\tau_j|\bar{q}_j)=\sum_{i \in r_j}\left\{ D_{ij}^2 \left(\begin{array}{cc}  \mathbf{x}_{ij} & w_{ij} \end{array}\right)^{T}
\var \left(\begin{array}{c} \hat{\boldsymbol{\beta}}_{\bar{q}_{j}} \\ \hat{\gamma}_{\bar{q}_{j}} \end{array}\right) \left(\begin{array}{cc}  \mathbf{x}_{ij} & w_{ij} \end{array}\right)\right\}
+\sum_{i \in r_j}D_{ij}^2 \var(y_{ij}).
\end{equation}

A first order approximation to $\var(\hat{\boldsymbol{\beta}}_{\bar{q}_{j}}, \hat{\gamma}_{\bar{q}_{j}})$ is obtained following \cite{Chambers2014} and \cite{bianchi:2015}. These approximated expressions lead to the following sandwich estimator:
\begin{equation}
 \widehat{\var} \left(\begin{array}{c} \hat{\boldsymbol{\beta}}_{\bar{q}_{j}} \\ \hat{\gamma}_{\bar{q}_{j}} \end{array}\right)={\scriptstyle \frac{n}{(n-p-1)}\frac{\sum_{j=1}^m \sum_{i \in s_j} \psi^2\left(\omega_{ij}^{-1}(y_{ij}-\mathbf{x}_{ij}^{T}\hat{\boldsymbol{\beta}}_{\bar{q}_{j}}- w_{ij} \hat{\gamma}_{\bar{q}_{j}})\right)}{\left\{\sum_{j=1}^m \sum_{i \in s_j}\psi^{\prime}\left(\omega_{ij}^{-1}(y_{ij}-\mathbf{x}_{ij}^{T}\hat{\boldsymbol{\beta}}_{\bar{q}_{j}}- w_{ij} \hat{\gamma}_{\bar{q}_{j}})\right)\right\}^{2}}\left(\left(\begin{array}{cc}  \tilde{\mathbf{X}} & \tilde{\mathbf{W}} \end{array}\right)^{T}\left(\begin{array}{cc}  \tilde{\mathbf{X}} & \tilde{\mathbf{W}} \end{array}\right)\right)^{-1}},
\end{equation}
where $\omega_{ij}$ is a robust estimator of the scale of the residual $y_{ij}-\mathbf{x}_{ij}^{T}\hat{\boldsymbol{\beta}}_{\bar{q}_{j}}- w_{ij} \hat{\gamma}_{\bar{q}_{j}}$ in area $j$. An estimator of the first-order approximation \eqref{prediction_var} is then

\begin{equation}\label{est_var}
\widehat{\var}(\hat{\tau}_{PATE_j}^{MQ}|\bar{q}_j)=\sum_{i \in r_j}\left\{ D_{ij}^2 \left(\begin{array}{cc}  \mathbf{x}_{ij} & w_{ij} \end{array}\right)^{T}
\widehat{\var} \left(\begin{array}{c} \hat{\boldsymbol{\beta}}_{\bar{q}_{j}} \\ \hat{\gamma}_{\bar{q}_{j}} \end{array}\right) \left(\begin{array}{cc}  \mathbf{x}_{ij} & w_{ij} \end{array}\right)\right\}
+\widehat{\var}(y_{ij})\sum_{i \in r_j}D_{ij}^2,
\end{equation}
where $\widehat{\var}(y_{ij})=(n-1)^{-1}\sum_{j=1}^m \sum_{i \in s_j} \left(y_{ij}-\mathbf{x}_{ij}^{T}\hat{\boldsymbol{\beta}}_{\bar{q}_{j}}- w_{ij} \hat{\gamma}_{\bar{q}_{j}}\right)^2$.

A corresponding estimator of the area-specific bias of the IPW-MQ estimator is
\begin{equation}\label{bias_MQ}
    \hat{B}(\hat{\tau}_{PATE_j}^{MQ}|\bar{q}_j)=\sum_{k=1}^m\sum_{i \in s_k}c_{ij}(\mathbf{x}_{ik}^{T}\hat{\boldsymbol{\beta}}_{\bar{q}_{k}}+ w_{ik} \hat{\gamma}_{\bar{q}_{k}})-\sum_{i \in \mathcal{U}_{j}}D_{ij}\left(\mathbf{x}_{ij}^{T}\hat{\boldsymbol{\beta}}_{\bar{q}_{j}}+ w_{ij} \hat{\gamma}_{\bar{q}_{j}}\right),
\end{equation}
where $c_{ij}=b_{ij}+D_{ij}I(i \in j)$ and 
\begin{displaymath}
\mathbf{b}_j=(b_ij)=\left( \sum_{i \in r_j}D_{ij}\left(\begin{array}{cc}  \mathbf{x}_{ij} & w_{ij} \end{array}\right) \right)\mathbf{W}_{\bar{q}_{j}}^{MQ}\left(\begin{array}{cc}  \tilde{\mathbf{X}} & \tilde{\mathbf{W}} \end{array}\right)\left( \left(\begin{array}{cc}  \tilde{\mathbf{X}} & \tilde{\mathbf{W}} \end{array}\right)^{T}\mathbf{W}_{\bar{q}_{j}}^{MQ} \left(\begin{array}{cc}  \tilde{\mathbf{X}} & \tilde{\mathbf{W}} \end{array}\right)\right)^{-1}.
\end{displaymath}
The final expression for the estimator of the MSE of IPW-MQ is just the sum of equation \eqref{est_var} and the square of equation \eqref{bias_MQ}:
\begin{equation}\label{MQ_predictor}
\widehat{MSE}(\hat{\tau}_{PATE_j}^{MQ}|\bar{q}_j)=\widehat{\var}(\hat{\tau}_{PATE_j}^{MQ}|\bar{q}_j)+\hat{B}^2(\hat{\tau}_{PATE_j}^{MQ}|\bar{q}_j).
\end{equation}

Following the approach of  \cite{bianchi:2015}, a further adjustment to the approximation of the MSE is needed to account for the variation due to the estimation of the area M-quantile coefficient $\bar{q}_j$ in the equation \eqref{MQ_predictor}. Therefore,
\begin{equation}\label{g4}
\var(\hat{\bar{q}}_j)=\left(\begin{array}{cc}  \tilde{\mathbf{X}}_j & \tilde{\mathbf{W}}_j \end{array}\right)\mathbf{G}_{\bar{q}_j}^{T}\mathbf{G}_{\bar{q}_j}\left(\begin{array}{cc}  \tilde{\mathbf{X}}_j & \tilde{\mathbf{W}}_j \end{array}\right)^{T} v^2_{\hat{\bar{q}}_j},
\end{equation}
where $\mathbf{G}_{\bar{q}_j}=n^{-1}\sum_{j=1}^{m}\left(\mathbf{H}^{-1}_{j \bar{q}_j}\left \{\partial_{\bar{q}_j} \mathbf{L}_{j \bar{q}_j}-  \partial_{\bar{q}_j} \mathbf{H}_{j \bar{q}_j} \mathbf{H}_{j \bar{q}_j}^{-1} \mathbf{L}_{j \bar{q}_j} \right \}   \right)$ with  $\mathbf{H}_{j \bar{q}_j}=\tilde{\mathbf{X}}_j^{T}\mathbf{W}_{\bar{q}_j}^{MQ} \tilde{\mathbf{X}}_j$, $\mathbf{L}_{j \bar{q}_j}=\tilde{\mathbf{X}}_j^{T}\mathbf{W}_{\bar{q}_j}^{MQ} \tilde{\mathbf{y}}_{j}$, $\partial_{\bar{q}_j} \mathbf{H}_{j \bar{q}_j}=\tilde{\mathbf{X}}_{j}^{T}\partial_{\bar{q}_j}\mathbf{W}_{j \bar{q}_j}^{MQ} \tilde{\mathbf{X}}_{j}$,$\partial_{\bar{q}_j}\mathbf{L}_{j  \bar{q}_j}=\tilde{\mathbf{X}}_j^{T}\partial_{\bar{q}_j}\mathbf{W}_{j \bar{q}_j}^{MQ} \tilde{\mathbf{y}}_{j}$, \linebreak
$\partial_{\bar{q}_j}\mathbf{W}_{j \bar{q}_j}^{MQ} =2 \mathbf{\Omega}_j \Big|\psi \left \{ \mathbf{\Omega}_j^{-1}(\tilde{\mathbf{y}}_{j}-\tilde{\mathbf{X}}_{j}^{T}\boldsymbol{\beta}_{\bar{q}_j}  \right\}\Big|\left \{\tilde{\mathbf{y}}_{j}-\tilde{\mathbf{X}}_{j}^{T}\boldsymbol{\beta}_{\bar{q}_j}  \right\}^{-1}$, $\Omega_j=diag(\omega_{ij}),~i \in s_j$ and $v^2_{\hat{\bar{q}}_j}=n_{j}^{-1}\sum_{i=1}^{n_j}(\hat{q}_{ij}-\hat{\bar{q}}_j)^2$ where $\hat{q}_{ij}$ are the M-quantile coefficients at unit level. This expression \eqref{g4} can be estimated by
\begin{equation}\label{est_g4}
\widehat{\var}(\hat{\bar{q}}_j)=\left(\begin{array}{cc}  \tilde{\mathbf{X}}_j & \tilde{\mathbf{W}}_j \end{array}\right)\hat{\mathbf{G}}_{\bar{q}_j}^{T}\hat{\mathbf{G}}_{\bar{q}_j}\left(\begin{array}{cc}  \tilde{\mathbf{X}}_j & \tilde{\mathbf{W}}_j \end{array}\right)^{T} \hat{v}^2_{\hat{\bar{q}}_j},
\end{equation}
The final form of the MSE estimator of $\hat{\tau}_{PATE_j}^{MQ}$ is then
\begin{equation}\label{MQ_predictor_final}
mse(\hat{\tau}_{PATE_j}^{MQ})=\widehat{\var}(\hat{\tau}_{PATE_j}^{MQ})+\hat{B}^2(\hat{\tau}_{PATE_j}^{MQ})+\widehat{\var}(\hat{\bar{q}}_j).
\end{equation}

The validity of model-based inference depends on the validity of the model assumed. We empirically evaluate the properties of small area predictors and corresponding
MSE estimators. In particular, we use Monte Carlo simulation to evaluate the performance of the proposed small area estimators and their
corresponding MSEs in comparison with the performance of the IPW-Direct estimator at small area level. Due to space constraints, the results and the discussion are not reported in the manuscript but they can be found in Supplementary Material, Section \ref{S-sec:supp_Simulation}. These results show that the proposed small area predictors, IPW-EBLUP and IPW-MQ, are much more efficient than the IPW-Direct and this suggests that it may be good to use these predictors to estimate the average treatment effect when sample size in each area becomes small.

In addition the benchmarking properties of the estimators are shown in Section \ref{S-sec:prop} of Supplementary Material.

\section{A design-based simulation based on real data}\label{sec:Application}
In this section we perform a design based simulation study using the 2015 Italian module of the EU-SILC  survey. The focus is on estimating the effect of permanent versus temporary contracts on the economic insecurity of households in different regions of Italy. This is one of the EU-SILC target variables in the domain of social exclusion/non-monetary household deprivation indicators.

In the design based simulation we consider the following substantive policy issue. Suppose policy makers are interested in evaluating the impact of temporary employment contracts on the economic insecurity  of households, measured by subjective poverty as  defind in  \cite{Kapteyn1988}, with potential consequences on consumption behaviour, life satisfaction and well-being in general.  
The increase in non-standard forms of employment in many countries appears to have contributed to rising in-work poverty \citep{Euro17,Cret2013}.  The development of forms of flexible employment may have both positive and negative consequences.  On the one hand it is expected to increase employment and reduce unemployment. On the other hand, this is often associated with greater economic insecurity and poorer working conditions.  
Relatively little research has been dedicated to the link between job instability and subjective poverty . According to some scholars, temporary as opposed to permanent employment contributes to lower general life satisfaction and well-being and a worse perceived household income situation. \cite{Scherer2009} investigates the social consequences of insecure employment (fixed-term contracts), taking into account information on current family life, future family plans and general well-being. The analysis, for Western European countries, confirms that insecure employment is accompanied by more problematic social and family situations. These negative consequences are partly shaped by the specific institutional context (welfare state and labour market conditions).
\cite{Filandri2018}, using the 2014 Italian wave module of the EU-SILC (EU Statistics on Income and Living Conditions) survey, find that subjective poverty is associated with instability of household members’ job contracts, with effects on other life domains, such as well-being, adequate level of consumption, social integration.

Differently from the previous literature, in this paper we adopt a causal perspective and consider the effect of temporary employment on the feeling about the household economic status. As discussed above, an overall negative effect of temporary employment is expected compared to permanent employment.  However, we expect the effect to be heterogeneous across Italian regions due to different quality, and cost of living.  The effects may be confounded by local institutional contexts (local welfare policies and labour market conditions), in addition to socio-demographics characteristics and information on the employment situation (work intensity and the skill level of the occupation). The presence of numerous regions with a very small sample size makes it difficult to obtain reliable direct estimates at the area level and motivates the use of SAE techniques. In this simulation each region of Italy is considered as a small area.



In this setting the units of the analysis are the Italian households and the treatment, Job stability, is a dichotomous variable that gets the value $1$ if the head of the household (or the household respondent) has a temporary job and $0$ if she/he has a permanent job at the time of the interview. We assume the existence of a causal path from this variable to the lowest monthly income to make ends meet, which is a subjective measure of the household economic status.
This is one of the EU-SILC target variables in the domain of social exclusion/non-monetary household deprivation indicators. Respondents are asked to provide their own assessed indication of the very lowest net monthly income that the household would have to get in order to make ends meet, that is, to pay its usual necessary expenses. We use this continuous outcome as a proxy variable for subjective poverty in the following analysis.  For the outcome model, as it is common for highly right skewed outcome distributions, we use the log transformation and then we consider the transformed values per individual in the household by dividing the total value by the equivalised household size.  We consider two sets of plausible confounders and predictors at the individual and household level. The individual characteristics concerns the head or the responsible person in the household. We assume unconfoundedness conditioning on the following set
of covariates: \texttt{Age}, \texttt{Gender}, \texttt{Education}, \texttt{Marital status}, \texttt{Tenure}, \texttt{Family type}. Conditioned on this set of covariates is necessary for the unconfoundedness assumption to hold. In addition, the \texttt{Number of rooms} in the house, the \texttt{Dwelling type}, the existence of problems related to  crime, violence and vandalism ,  in the local area from the point of view of the respondents (\texttt{Crime}), and the \texttt{Household disposable income} are used as additional predictors of the outcome. 

The aim of the design-based simulation is to compare the performance of different estimators for the impact in each domain under repeated sampling from a fixed population. For this reason we consider the sample of the workforce, aged between 25 and 80, in the 20 administrative regions of Italy, based on the 2015 Italian module of the EU-SILC survey as a pseudo-population (population hereafter). Due to sample size requirements Abruzzo and Molise are aggregated, leading to 19 areas. After accounting for common support within all areas $11011$ units are left, from which $1254$ units belong to the treated and the rest to the control group. The area population sizes range from 152 to 1329 with an average of 580. Figure \ref{S-fig:com-supl} in Section \ref{S-sec:supp_app} of the Supplementary Material shows the overall common support of the propensity scores among treated and control groups. Balancing the covariates within each area has been verified by running t-tests on the difference in the average value of the propensity scores by treatment status. Table \ref{S-tab:balance} in Section \ref{S-sec:supp_app} shows that all covariates are balanced in this population for all 19 areas. 

The original estimates of the impacts are considered as the true $\tau_{j}$ parameters of the population level. This pseudo-population is then kept fixed over the Monte Carlo simulations. We draw $S=1000$ independent random samples without replacement this balanced population with regional common support, by randomly selecting individuals in the 19 regions with sample sizes in each areas set to $10\%$ of the its population size (resulting in a proportional stratified sampling).
The samples from each region are drawn not considering the treatment status. This means that the sample of a specific region might include both treated and control units, or it might only contain the observations from one of the two groups. This is to show one of the advantages of using our proposed method, because, as previously stated, our methods can be used even if the sample size of the treated or controlled group is zero in the domain of interest. Three different estimators are evaluated in this simulation study: the IPW-Direct (\ref{eq:SATE3}), the IPW-EBLUP (see Section \ref{subsec:EBLUP-est}) and the IPW-MQ (see Section \ref{subsec: MQ_est}). For the estimator based on M-quantile approach the influence function is the Huber-type function with tuning constant equal to $1.345$ for the continuous response and $1.6$ for the binary variable in the propensity scores estimation \citep{chambers2006,Chambers2016}. For each estimator and for each small area, we computed the Monte Carlo estimate of the percentage of relative bias and the percentage of relative root MSE and the corresponding efficiency. 

\begin{figure}[H]
    \centering
    \includegraphics[width=0.9\linewidth]{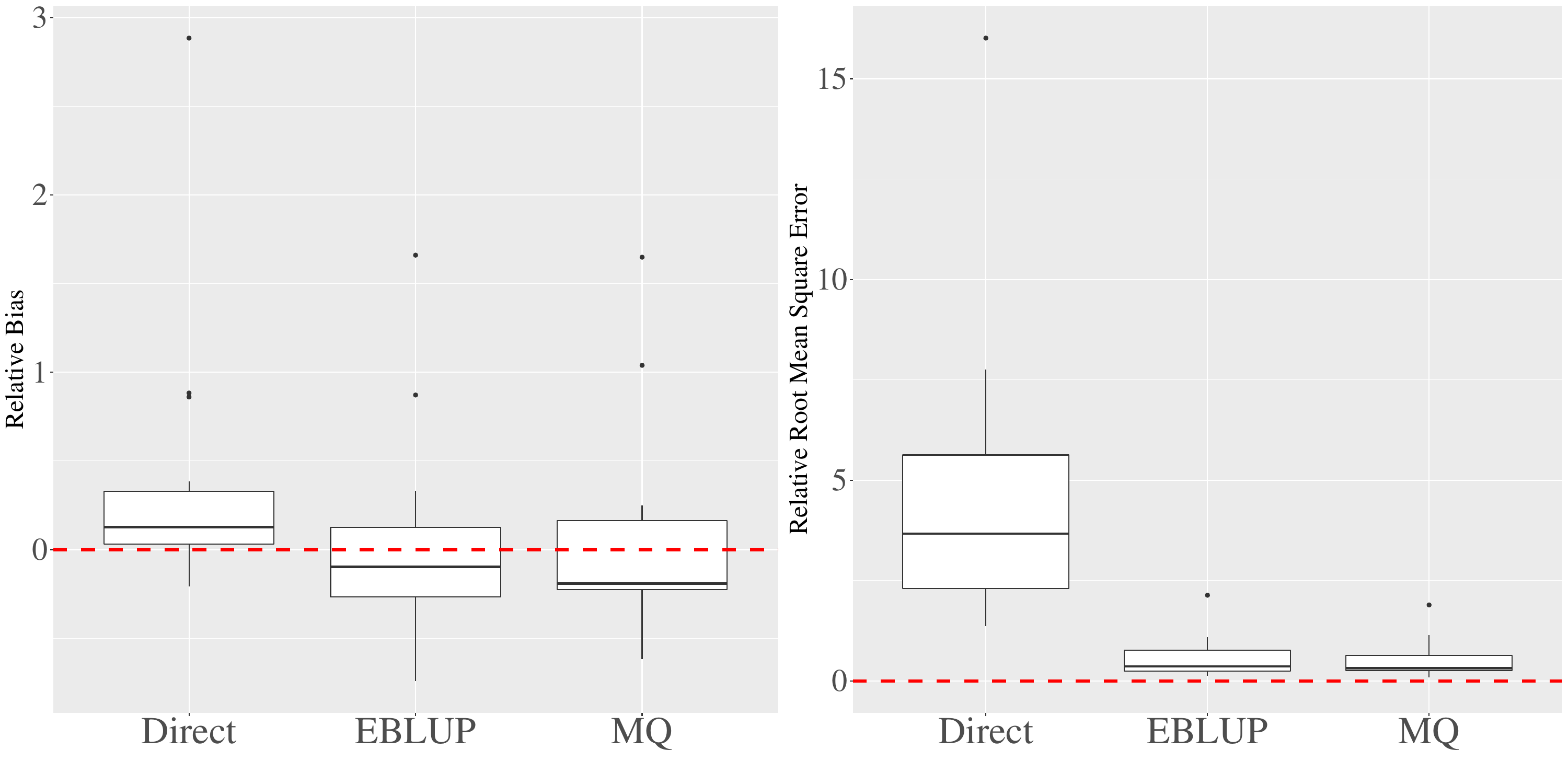} 
    \caption{Boxplots of the median values of area-specific relative bias and relative root mean square error computed over 1000 replications. Note that the Direct, EBLUP and MQ stand for IPW-Direct, IPW-EBLUP and IPW-MQ, respectively.}
    \label{fig:app_rrmse}
\end{figure}

Figure \ref{fig:app_rrmse} illustrates the box plots of the median values of area-specific relative bias and relative root MSE computed over replications, confirming the characteristics of the different estimators. We see that IPW-EBLUP and IPW-MQ work well in terms of both bias and relative root MSE compared with the IPW-Direct. This point is also highlighted in the series of model-based simulation studies in Section \ref{S-sec:supp_Simulation} of the Supplementary Material. Figure \ref{fig:het_est2} shows that IPW-MQ and IPW-EBLUP  outperform the IPW-Direct in capturing the heterogeneity of the average treatment effects over the areas. It illustrates that the distribution of the estimated effects by IPW-MQ and IPW-EBLUP (solid blue and red liens) is closer to the true distribution of the effects (dashed line) than the one estimated by IPW-Direct. The direct estimator is not even able to cover the whole rage of the true range of the treatment effects.

\begin{figure}[H]
    \centering
    \includegraphics[width=0.9\linewidth]{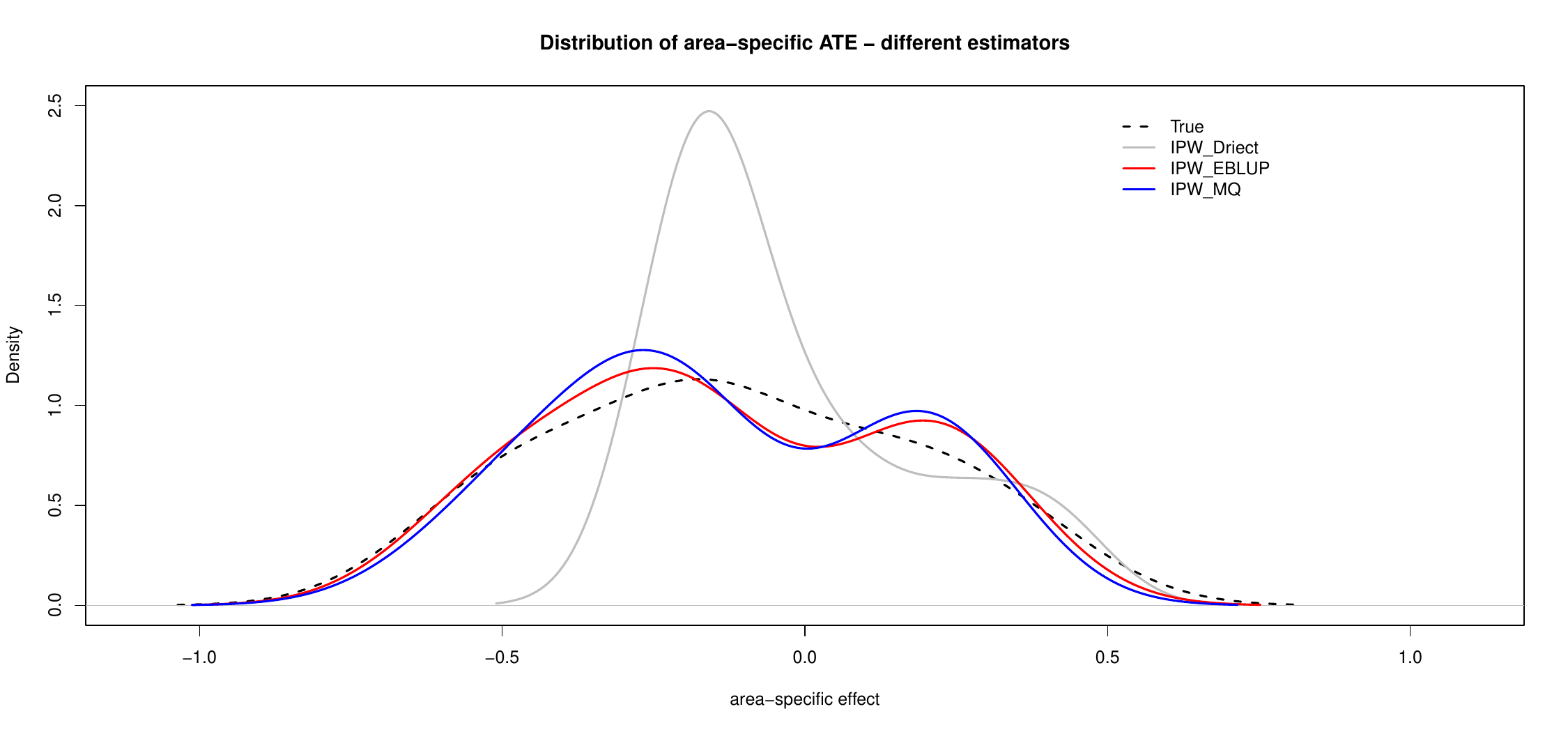}
    \caption{Performance of different estimators in capturing the distribution of heterogeneous effects across areas.}
    \label{fig:het_est2}
\end{figure}

The relative efficiencies of the proposed estimators with respect to IPW-Direct are computed as the
ratio of the average actual MSE for each area to the average actual MSE of the IPW-Direct.  Table \ref{tab:app_eff} presents the summary statistics over the 19 regions in the study. A value less than 100 for this ratio indicates that the MSE of the model-based
estimate (i.e. IPW-EBLUP, IPW-MQ )is smaller than that of the direct estimate. The results reported in Table \ref{tab:app_eff} indicate that the best method for this data appears to be the robust version, IPW-MQ. These results are consistent for all the areas in the study.

\begin{table}[H]
\centering
\caption{The efficiency of each estimator compared to IPW-Direct. Summary statistics over 19 regions in the study.}\label{tab:app_eff}
\vspace{0.5cm}
\begin{tabular}{lrrrrr}
  \hline
Method& Min. & 1st Qu. & Median  & 3rd Qu. & Max. \\ 
  \hline
IPW-EBLUP &25.73 & 40.09 & 47.25 &  52.11 & 61.39 \\ 
  IPW-MQ & 24.03 & 39.22 & 44.92 &  50.31 & 59.01 \\ 
   \hline
\end{tabular}
\end{table}

\begin{figure}
	\centering
	\includegraphics[width=1\linewidth]{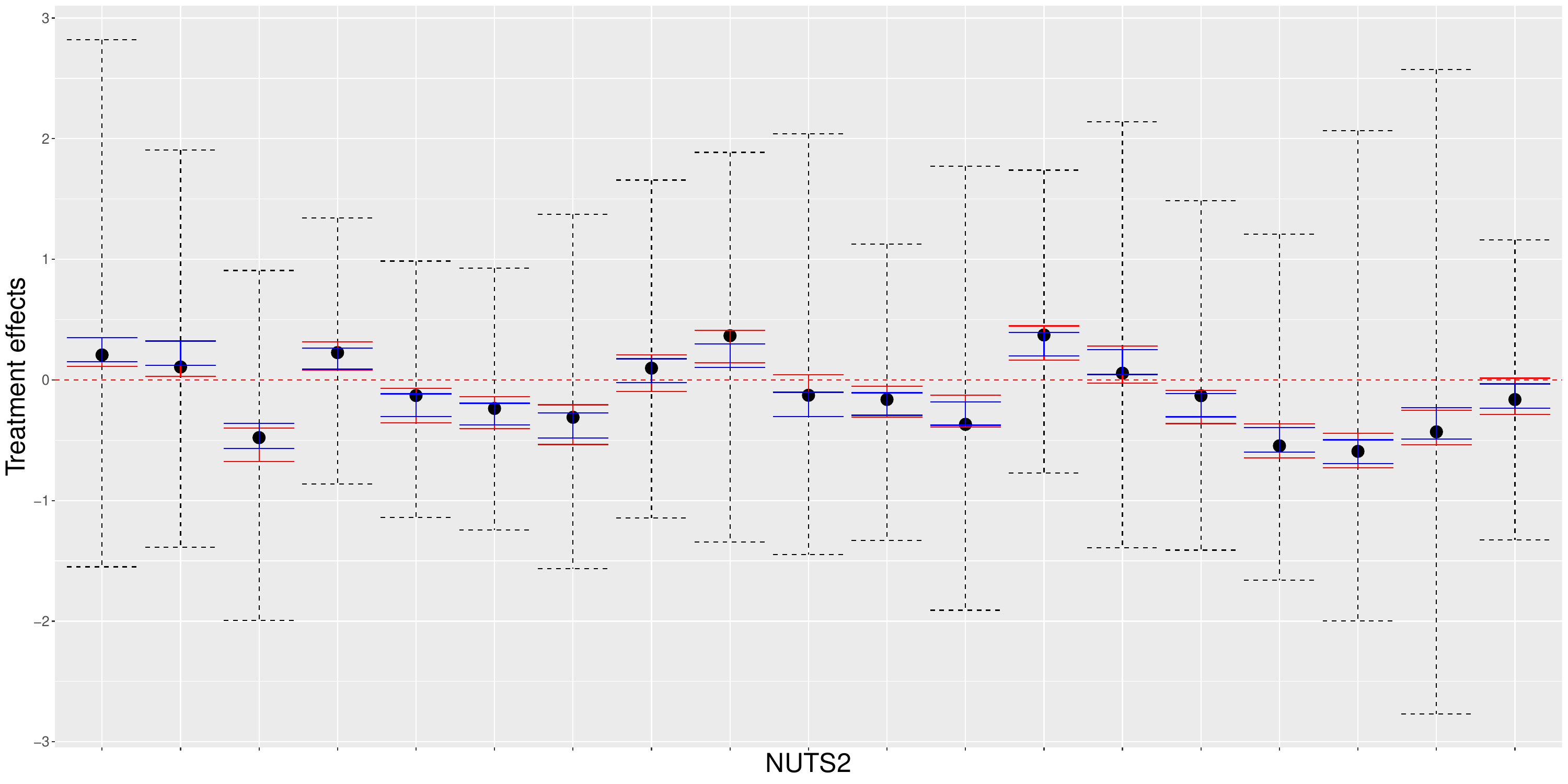}
	\caption{95\% confidence intervals (CI) based on the quantiles of 1000 Monte Carlo replication. True values of the effect for the 19 NUTS2 area are depicted by black dots. CIs are shown for the Direct estimator in dashed black , for IPW-EBLUP in solid red and for IPW-MQ in solid blue line.}
	\label{fig:CI_compare}
\end{figure}

Figure \ref{fig:CI_compare} illustrate the 95 \% Confidence intervals that are obtained by using the quantiles of the 1000 simulation estimates. In this illustration we can see that the length of the intervals for IPW-Direct estimator is much larger than our proposed IPW-EBLUP and IPW-MQ estimators due to the large variance of this estimator. This leads to the point that for all the 19 regions the CI of the direct estimator contain zero, implying that the direct method cannot not identify any significant effect and does not distinguish the heterogeneity of the effects among different areas.  On the contrary, the CIs for IPW-EBLUP and IPW-MQ  only contain zero in cases where the true area effects are very close to the zero line. Although the length of the intervals for our estimators are considerably lower than the direct estimator, they still mostly encompass the true values and manage the capture the heterogeneous effects among different regions.


\section{Conclusion}\label{sec:conclusion}
Small area techniques provide the official statistics for politicians and decision makers using sample surveys and other sources of information. However to the extent of our knowledge there is no link between this literature and that on causal inference, even though sometimes the statements in the former literature are interpreted in a causal way. 

In this paper we propose a methodological framework that links the two streams of literature and emphasise the relevance of such methods in many applications to real data. Our proposed methods take account of the heterogeneity of the effects across areas even at a very fine level (small area level). This allows policy makers and decision takers to know the impact of a given policy for a finer geographic, socio-demographic, or socio-economic grid and, consequently, to plan better local-targeted interventions. 

Some of the usual assumptions for making causal inference with observational data are revisited and modified to be consistent with the context of small area estimations. 
The proposed methods IPW-EBLUP and IPW-MQ are mainly based on weighting with propensity scores. These estimators inherit the properties of a doubly robust estimators, since both a model to estimate the scores and another model to predict the outcome are used for the  part of the population that is not observed. This means that if one of these two models is misspecified the estimator is still consistent.

For each of the proposed estimators, IPW-EBLUP and IPW-MQ, we developed an analytical MSE estimator under the assumption that the propensity score is known. We also suggest a correction for the bias in the analytical MSE, that can occur due to the estimation of the propensity score, by proposing two different bootstrap methods, defined as a parametric bootstrap and modified random effect block bootstrap, for IPW-EBLUP and IPW-MQ, respectively. The performance of the MSE estimators is studied via simulations. 

Monte Carlo model based simulations are used to evaluate the performance of the proposed estimators in comparison with the performance of the IPW-Direct at small area levels. The results show that the proposed small area predictors, IPW-EBLUP and IPW-MQ, are much more efficient than the IPW-Direct and this suggests that it may be best to use these predictors to estimate the average treatment effect when the sample size in each area becomes small. However, as expected, these methods manifest higher bias than the direct estimator.  
 
The application to real data, even if conducted as a design-based simulation analysis,  has shown the potential of the proposed method in reconstructing the detail of the impact at the regional level, albeit with differences in the performance of the estimators. Job stability affects the perception of economic insecurity, but not in a homogeneous way in the different regions. The effect is negative in most cases with even significant differences, which we can attribute to the different levels of quality and cost of living, as well as to a different social context in general. Once again, this highlights the importance of adopting local policies to support families and combat poverty.

As future lines of research, we plan to extend our results for other robust estimators, such as REBLUP \citep{sinha:2009}. Moreover, due to the presence of bias in IPW-EBLUP and IPW-MQ, observed in our simulation experiment, we would like to investigate bias calibration methods to make the approach predictive rather than projective. Finally, we aim to exploit the use of other matching techniques by properly defining distance measures and including the predicted random effects in the matching algorithm in small area estimation.

\bibliographystyle{chicago}
\bibliography{CISAE}

\end{document}


\def\spacingset#1{\renewcommand{\baselinestretch}%
	{#1}\small\normalsize} \spacingset{1}

\begin{center}
{\large\bf SUPPLEMENTARY MATERIAL}
\end{center}
\bigskip

\spacingset{1.45} 
\section{Double robust properties} \label{sec:doub_robust}
When in equation \eqref{P-eq:PATE3} the weights are not re-normalized the estimator assumes the form:

\begin{align} \label{eq:PATE3Al}
\tau_{PATE_{j}}^{\star} & =\left. \left(\sum_{i\in s_{j}}\left[\frac{w_{ij}y_{ij}}{\hat{e}(\mathbf{x}_{ij})}\right]+\sum_{i\in r_{j}}\left[\frac{w_{ij}\hat{y}_{ij}}{\hat{e}(\mathbf{x}_{ij})}\right]\right) N_{j}^{-1} \right. - \nonumber \\
&
\left. \left(\sum_{i\in s_{j}}\left[\frac{(1-w_{ij})y_{ij}}{1-\hat{e}(\mathbf{x}_{ij})}\right]+\sum_{i\in r_{j}}\left[\frac{(1-w_{ij})\hat{y}_{ij}}{1-\hat{e}(\mathbf{x}_{ij})}\right]\right)N_j^{-1}\right.. 
\end{align}
 But, unfortunately, this estimator is no more double robust nor consistent and an adjustment term is needed to render the desirable properties. To meet these properties two components have to be added to the estimator \eqref{eq:PATE3Al} following \cite{Robins:1994}:
 \begin{align} \label{eq:PATE3AlDR}
\hat{\tau}_{PATE_{j}}^{\star DR} & =\left. \left(\sum_{i\in s_{j}}\left[\frac{w_{ij}y_{ij}}{\hat{e}(\mathbf{x}_{ij})}\right]+\sum_{i\in r_{j}}\left[\frac{w_{ij}\hat{y}_{ij}}{\hat{e}(\mathbf{x}_{ij})}\right]-\sum_{i\in U_{j}}\left[\frac{(w_{ij}-\hat{e}(\mathbf{x}_{ij}))\hat{y}_{ij}}{\hat{e}(\mathbf{x}_{ij})}\right]\right) N_{j}^{-1} \right. - \nonumber \\
&
\left. \left(\sum_{i\in s_{j}}\left[\frac{(1-w_{ij})y_{ij}}{1-\hat{e}(\mathbf{x}_{ij})}\right]+\sum_{i\in r_{j}}\left[\frac{(1-w_{ij})\hat{y}_{ij}}{1-\hat{e}(\mathbf{x}_{ij})}\right]+\sum_{i\in U_{j}}\left[\frac{(w_{ij}-\hat{e}(\mathbf{x}_{ij}))\hat{y}_{ij}}{1-\hat{e}(\mathbf{x}_{ij})}\right]\right)N_j^{-1}\right. ,
\end{align}
where $DR$ stands for double robust. 
We have compared the two estimators by simulation experiments and estimator \eqref{P-eq:PATE3} is more efficient than \ref{eq:PATE3AlDR}. The results are not reported in the paper for reason of space, but are available from the authors upon request.

\section{Proof of Theorem 1}\label{sec_proofth}
\begin{proof}
	The starting point is:
	\begin{equation*} \label{eq:PATE_final1}
	\hat{\tau}_{PATE_{j}} =\left(\sum_{i\in s_{j}}a_{ij}y_{ij}+\sum_{i\in r_{j}}a_{ij}\hat{y}_{ij}\right) \left(\sum_{i=1}^{N_{j}} a_{ij}\right)^{-1} - \left(\sum_{i\in s_{j}}b_{ij}y_{ij}+\sum_{i\in r_{j}}b_{ij}\hat{y}_{ij}\right) \left(\sum_{i=1}^{N_{j}} b_{ij}\right)^{-1},
	\end{equation*}
	The first component of the estimator can be written as:
	\begin{equation*} \label{eq:exp1}
	\left(\sum_{i\in s_{j}}a_{ij}y_{ij}+\sum_{i\in r_{j}}a_{ij}\hat{y}_{ij}\right) A_{N_j}^{-1}.
	\end{equation*}
	The values $\pm \sum_{i\in U_{j}}a_{ij}y_{ij}$ and $\pm \sum_{i\in s_{j}}a_{ij}\hat{y}_{ij}$ can be added and subtracted to the previous expression:
	\begin{equation*} \label{DR-proof1}
	\left(\sum_{i\in U_{j}}a_{ij}y_{ij}-\sum_{i\in U_{j}}a_{ij}y_{ij}+\sum_{i\in s_{j}}a_{ij}\hat{y}_{ij}-\sum_{i\in s_{j}}a_{ij}\hat{y}_{ij}+\sum_{i\in s_{j}}a_{ij}y_{ij}+\sum_{i\in r_{j}}a_{ij}\hat{y}_{ij}\right)  A_{N_j}^{-1}.
	\end{equation*}
	Here $\sum_{i\in s_{j}}a_{ij}\hat{y}_{ij}+\sum_{i\in r_{j}}a_{ij}\hat{y}_{ij}=\sum_{i\in U_{j}}a_{ij}\hat{y}_{ij}$, then 
	\begin{align*} \label{DR-proof2}
	& \left(\sum_{i\in U_{j}}a_{ij}y_{ij}-\sum_{i\in U_{j}}a_{ij}y_{ij}+\sum_{i\in U_{j}}a_{ij}\hat{y}_{ij}-\sum_{i\in s_{j}}a_{ij}\hat{y}_{ij}+\sum_{i\in s_{j}}a_{ij}y_{ij}\right) A_{N_j}^{-1}\\
	=& \left(\sum_{i\in U_{j}}a_{ij}y_{ij}^1-\sum_{i\in U_{j}}a_{ij}(y_{ij}-\hat{y}_{ij})+\sum_{i\in s_{j}}a_{ij}(y_{ij}-\hat{y}_{ij})\right)  A_{N_j}^{-1}\\
	=& A_{N_j}^{-1}\sum_{i\in U_{j}}a_{ij}y_{ij}^1-A_{N_j}^{-1} \sum_{i\in U_{j}}a_{ij}(y_{ij}-\hat{y}_{ij})+A_{N_j}^{-1}\sum_{i\in s_{j}}a_{ij}(y_{ij}-\hat{y}_{ij})\\
	= & A_{N_j}^{-1}\sum_{i\in U_{j}}a_{ij}y_{ij}^1-A_{N_j}^{-1}\sum_{i\in U_{j}}a_{ij}(y_{ij}-\hat{y}_{ij}),
	\end{align*}
	where since ${y}_{ij}= w_{ij}{y}^{1}_{ij}+(1-w_{ij}){y}^{0}_{ij}$, then it can be written that $ w_{ij}{y}_{ij}=w_{ij}{y}^{1}_{ij}=$ and $(1-w_{ij}){y}_{ij}=(1-w_{ij}){y}^{0}_{ij}$; the value $A_{N_j}^{-1}\sum_{i\in s_{j}}a_{ij}(y_{ij}-\hat{y}_{ij})$ is negligible because the quantity $a_{ij}E[(y_{ij}-\hat{y}_{ij})]$ is computed respect the sampled unit and $A_{N_j} \rightarrow \infty$. So the last term of the above expression disandixppears.
	
	The same approach is used for the second component of the estimator. In particular,
	\begin{equation*} \label{eq:exp2}
	\left(\sum_{i\in s_{j}}b_{ij}y_{ij}+\sum_{i\in r_{j}}b_{ij}\hat{y}_{ij}\right) B_{N_j}^{-1}.
	\end{equation*}
	
	Then in the second term of the estimator the values $\pm \sum_{i\in U_{j}}b_{ij}y_{ij}$ and $\pm \sum_{i\in s_{j}}b_{ij}\hat{y}_{ij}$ are added and subtracted:
	\begin{align*} \label{DR-proof1b}
	& \left(\sum_{i\in U_{j}}b_{ij}y_{ij}-\sum_{i\in U_{j}}b_{ij}y_{ij}+\sum_{i\in U_{j}}b_{ij}\hat{y}_{ij}-\sum_{i\in s_{j}}b_{ij}\hat{y}_{ij}+\sum_{i\in s_{j}}b_{ij}y_{ij}\right) B_{N_j}^{-1}\\
	= & B_{N_j}^{-1}\sum_{i\in U_{j}}b_{ij}y_{ij}^0-B_{N_j}^{-1} \sum_{i\in U_{j}}b_{ij}(y_{ij}-\hat{y}_{ij})+B_{N_j}^{-1}\sum_{i\in s_{j}}b_{ij}(y_{ij}-\hat{y}_{ij}).
	\end{align*}
	As before the last term is negligible, then the previous expression can be written as
	     \begin{equation*} \label{DR-proof5}
	B_{N_j}^{-1}\sum_{i\in U_{j}}b_{ij}y_{ij}^0-B_{N_j}^{-1}\sum_{i\in U_{j}}b_{ij}(y_{ij}-\hat{y}_{ij}).
	\end{equation*} 	
	
	Using the results from the paper by \cite{Chandra1996} on the Strong Law of Large Numbers (SLLN) for weighted average estimator the proposed PATE estimator is double robust:
	
		$$ A_{N_j}^{-1}\sum_{i\in U_{j}}a_{ij}y_{ij}^1-A_{N_j}^{-1}\sum_{i\in U_{j}}a_{ij}(y_{ij}-\hat{y}_{ij}) -E_j\left[Y_{ij}^{1}\right]+ E_j\left[Y_{ij}-\hat{Y}_{ij}\right] \overset{a.s.}{\to} 0 ,$$
	
	and 
		
	$$B_{N_j}^{-1}\sum_{i\in U_{j}}b_{ij}y_{ij}^0-B_{N_j}^{-1}\sum_{i\in U_{j}}b_{ij}(y_{ij}-\hat{y}_{ij}) -E_j\left[Y^{0}_{ij}\right]+E_j[{Y}_{ij}-\hat{Y}_{ij}] \overset{a.s.}{\to} 0 , $$
	
	where a.s. means almost surely. Therefore:

	$$Pr\left(\lim\limits_{N_{j}\rightarrow \infty}\hat{\tau}_{PATE_{j}}= E_j\left[Y^{1}_{ij}\right]-E_j\left[{Y}_{ij}-\hat{Y}_{ij}\right]-E_j\left[Y^{0}_{ij}\right]+E_{j}\left[{Y}_{ij}-\hat{Y}_{ij}\right] =E_{j}\left[Y^{1}_{ij}\right]- E_{j}\left[Y^{0}_{ij}\right]\right)=1.$$

	In addition almost surely convergence implies convergence in probability \citep[p.23]{Jiang2010}, therefore:
	$$ \hat{\tau}_{PATE_{j}} \overset{p}{\to}\tau_{PATE_{j}} \qquad \text{as} \qquad N_{j}\rightarrow \infty, $$
	
	\noindent which means  $\hat{\tau}_{PATE_{j}}$ is a consistent estimator of $\tau_{PATE_{j}}$.
\end{proof}

\section{ Asymptotic properties of the IPW-EBLUP}\label{sec:asym:eblup}

\begin{proof}
According to the Theorem \ref{P-theo:doub-rob}, IPW-EBLUP is double robust and consistent since it is special case of estimator in \eqref{P-eq:PATE3}. In addition,
this is an unbiased estimator of $\tau_{j}$. Note that $\hat{y}_{ij}$ is an unbiased empirical predictor of the $y_{ij}$, fitted using ML or REML and $\hat{e}(\mathbf{x}_{ij})$ are assumed to be deterministic for the reasons explained in the conditions of the Theorem \ref{P-theo:doub-rob} so starting from \eqref{P-eq:IPW2} we have :
$$ E_{j}\left[\hat{\tau}_{PATE_j}^{EBLUP}-\tau_j\right]= \sum_{i \in r_{j}}D_{ij}(E_{j}[\hat{y}_{ij}]-E_{j}[y_{ij}])=0$$
 
Now considering equations \eqref{P-eq:MSE1} and \eqref{P-eq:MSE2} according to \citet{Datta2000} :
\begin{equation*}
  MSE(\hat{\tau}_{PATE_{j}}^{EBLUP}) = \mathbf{D}_{j}^{T}\mathbf{\tilde{Z}}_{j}\boldsymbol{\Sigma}_{\omega}\left(\mathbf{I}-\mathbf{\tilde{Z}}^{T}\mathbf{V}^{-1}\mathbf{Z}\boldsymbol{\Sigma}_{\omega}\right)\mathbf{\tilde{Z}}_{j}^{T}\mathbf{D}_{j}+\mathbf{D}_{j}^{T} \mathbf{c}_{j}(\mathbf{\tilde{X}}^{T}\mathbf{V}^{-1}\mathbf{\tilde{X}})\mathbf{c}_j^{T}\mathbf{D}_{j} +tr\left(\mathcal{S} \mathbf{V}\mathcal{S}^{T} \mathcal{I}^{-1}\right) + o(m^{-1}).
\end{equation*}
The central limit theorem (CLT) follows directly then to achieve the asymptotic normality of the estimator. 
Since the $\hat{\tau}_{PATE_j}^{EBLUP}$ is unbiased the second-order correct estimator of the variance is $\hat{\mathscr{V}}_{j}(\theta)= mse(\hat{\tau}_{PATE_{j}}^{EBLUP})$ from equation \eqref{P-eq:mse-eblup}.

\end{proof}

\section{ Bias corrected MSE via bootstrap procedure}\label{sec:boots-both}
The proposed analytical MSE estimators (Section \ref{P-sec:MSE}) do not take into account the variability due to the estimation of the propensity scores. For adding this component of variability re-sampling techniques are proposed. In particular, for IPW-EBLUP we suggest using a parametric bootstrap technique, such as that proposed by \cite{Gonzalez-Manteiga:2008}, or a non-parametric bootstrap procedure as in \cite{opsomer2008}. For IPW-MQ, we suggest applying an outlier robust bootstrap approach that is a modified version of the block-bootstrap approach of \cite{Chambers:2013}.

\subsubsection*{Parametric bootstrap for IPW-EBLUP}
The use of a parametric bootstrap to get an estimate of the MSE is very common in SAE \cite{hall:2006b}. In this section modifications of existing techniques will allow us to capture part of the variation in the IPW-EBLUP that is due to the estimation of the propensity scores. The underlying assumption is that the error terms (area-specific and individual) are normally distributed.  

The steps of this extended parametric bootstrap are as follows:

\begin{enumerate}[label=\textbf{Step~\arabic*},leftmargin=*]
	\item Fit model \eqref{P-eq:est2} to the initial data obtaining estimated value $(\boldsymbol{\hat{\beta},\hat{\alpha}}) $ for the fixed part and $(\hat{\sigma}^{2}_{e}, \hat{\sigma}^{2}_{\gamma}, \hat{\sigma}^2_{u}, \hat{\sigma}^{2}_{\nu})$ for the random components.
	\item Construct the bootstrap vectors $\epsilon^{\ast}$ , $u^{\ast}$ , $ \gamma^{\ast} $ and $\nu{\ast}$ whose elements are independent copies of $N(0, \hat{\sigma}^{2}_{e})$, $ N(0,\hat{\sigma}^{2}_{\gamma})$,  $N(0,\hat{\sigma}^2_{u})$ and $N(0,\hat{\sigma}^{2}_{\nu})$, respectively, for the entire population.  
	\item Construct the bootstrap population data $(y_{ij}^{\ast},\mathbf{x}_{ij}, w^{\ast}_{ij})$ from the model 
	$$y_{ij}^{\ast}=\mathbf{\tilde{x}}_{ij}^{T}\boldsymbol{\tilde{\beta}}+w_{ij}^{\ast}\gamma_{j}^\ast+u_{j}^{\ast}+\epsilon_{ij}^{\ast},$$
	where $ w_{ij}^{\ast}\sim \operatorname{Bernoulli}(p_{ij}^{\ast})$ and $p_{ij}^{\ast}=\Lambda^{-1}(\mathbf{x}_{ij}^{T}\boldsymbol{\alpha}+\nu_{j}^{\ast}) $.
	\item Draw a sample from the bootstrap population of size $n_j$ for each area using the stratified random sampling and fit the models \eqref{P-eq:est2}, \eqref{P-eq:est3}. Then calculate the bootstrap estimated outcomes $\hat{y}_{ij}^{\ast}$ and propensity score $\hat{p}_{ij}^{\ast}$. 
	\item Calculate the bootstrap errors $\hat{\tau}_{j}^{\ast}-\tau_{j}^{\ast}$, for each $j=1, \cdots, m$.  The value $\tau_{j}^{\ast}$ is the true bootstrap value obtained by replacing $\hat{y}_{ij}^{\ast}$ and $p_{ij}^{\ast}$ in equation \eqref{P-eq:PATE3} and the $\hat{\tau}_{j}^{\ast}$ is the bootstrap estimate and it is derived by replacing the $\hat{y}_{ij}^{\ast}$ and $\hat{p}_{ij}^{\ast}$ in the same equations. 
	\item Repeat steps $2-5$ $B$ times. In the $b$th bootstrap replication, let $\tau_{j}^{\ast(b)}$ be the quantity of interest for area $j$ and $\hat{\tau}_{j}^{\ast(b)}$ its bootstrap estimate. A bootstrap estimator of area $j$ for the variability contribution of the estimates of propensity scores to the MSE of IPW-EBLUP can be calculated as
	\begin{equation}
	    \var(\hat{\tau}_{j})=\frac{1}{B}\sum_{b=1}^{B}(\hat{\tau}_{j}^{\ast(b)}-\tau_{j}^{\ast(b)})^2.
	\end{equation}
\end{enumerate} 
 This value has to be added to the estimator of the MSE in equation \eqref{P-eq:mse-eblup}.
 
\subsubsection*{Modified random effect block bootstrap method for IPW-MQ}
In this section we propose an M-quantile ensemble modelling approach to recreat the hierarchical population variability in the original population, combining it with a modified version of the block bootstrap method proposed by \cite{Chambers:2013} for the estimation of the variability contribution of the estimates of propensity scores to the MSE of the IPW-MQ predictor. The advantage of this method is that there are no assumptions on the underlying distributions of the error terms. 
 
The steps of the modified random effect block bootstrap are as follows:




\begin{enumerate}[label=\textbf{Step \arabic*},leftmargin=*]
	\item Compute the robust estimate $\hat{\boldsymbol{\beta}}_{0.5}$ for the fixed effects vector defining the linear regression M-quantile of order $q=0.5$ and calculate the marginal residuals defined as $\hat{r}_{ij}=y_{ij}-\mathbf{x}_{ij}^{T}\hat{\boldsymbol{\beta}}_{0.5}-w_{ij} \hat{\gamma}_{0.5}$. 
	\item Calculate the level two and level one empirical residuals generated by the
fitted linear quantile regression of order $q=0.5$ as:
	\begin{equation}\label{gamma_boot}
	\hat{\gamma}_{j}=\frac{\sum_{i=1}^{n_{j}}(w_{ij}-w_{.j})(\hat{r}_{ij}-\hat{r}_{.j})}{\sum_{i=1}^{n_{j}}(w_{ij}-w_{.j})^2},
	\end{equation}
	\begin{equation}
	\hat{u}_{j}=\hat{r}_{.j}-\hat{\gamma}_{j}w_{.j},
	\end{equation}
	\begin{equation}
	\hat{\epsilon}_{ij}=\hat{r}_{ij}-\hat{u}_{j}-w_{ij}\hat{\gamma}_{j},
	\end{equation}
	where $w_{.j}=\frac{1}{n_{j}}\sum_{i=1}^{n_{j}}w_{ij}$ and $\hat{r}_{.j}=\frac{1}{n_{j}}\sum_{i=1}^{n_{j}}\hat{r}_{ij}$.
	\item Calculate the moment-based estimates $\hat{\theta}=(\hat{\sigma}_\gamma^{2},\hat{\sigma}_{u}^{2},\hat{\sigma}_{\epsilon}^{2})$ of the between area and
within area variance components defined by the hierarchical linear model \eqref{P-model_mixed} (e.g., using the \texttt{mbest} package within \texttt{R}) and then center around zero and re-scale the empirical residuals and random effects calculated in the previous step to achieve consistency:
	$$\hat{\gamma}_{j}^{cs}=\hat{\sigma}_\gamma \hat{\gamma}_{j}^{c}\left(\frac{1}{m}\sum_{h=1}^{m}(\hat{\gamma}_{h}^{c})^{2}\right)^{(-1/2)},$$
	$$\hat{u}_{j}^{cs}=\hat{\sigma}_{u}\hat{u}_{j}^{c}\left(\sum_{h=1}^{m}(\hat{u}_{h}^{c})^{2}\right)^{(-1/2)},$$
	$$\hat{\epsilon}_{ij}^{cs}=\hat{\sigma}_{\epsilon}\hat{\epsilon}_{ij}\left(\frac{1}{n}\sum_{h=1}^{m}\sum_{k=1}^{n_{h}}(\hat{\epsilon}_{kh}^{c})^{2}\right)^{(-1/2)},$$
where $\hat{\gamma}_{j}^{c}=\hat{\gamma}_{j}-\frac{1}{m}\sum_{h=1}^{m}\hat{\gamma}_{h}$, $\hat{u}_{j}^{c}=\hat{u}_{j}-\frac{1}{m}\sum_{h=1}^{m}\hat{u}_{h} $ and $\hat{\epsilon}_{ij}^{c}=\hat{\epsilon}_{ij}^{c}-\frac{1}{n}\sum_{h=1}^{m}\sum_{k=1}^{n_{h}}\hat{\epsilon}_{kh}$.
	\item At this step we propose a re-sampling scheme that is mainly plausible if the mutual independence between the random components hold, that is, $u_{j}$,$\gamma_{j}$ and $\epsilon_{ij}$ are mutually independent. Therefore, in the estimation procedure the model is fit by imposing the restriction that covariance between random slopes and random intercepts is zero. That can avoid  the over-parametrisation of the model. That is, for each bootstrap iteration $b$,
	\begin{enumerate}
		\item Generate the random slopes values for the $m$ areas by drawing a simple random sample of size $m$ with replacement ($srswr$) from the vector $\hat{\boldsymbol{\gamma}}^{sc}=(\hat{\gamma}_{1}^{sc},\cdots \hat{\gamma}_{m}^{sc})$. Denote this sample    by $\boldsymbol{\gamma}^{\ast}$.
		\item Similarly, generate a vector of random intercepts $\hat{\boldsymbol{u}}^{\ast}$ for $m$ areas by independently drawing simple random samples of size $m$ with  replacement from the vector $\hat{\boldsymbol{u}}^{sc}=(\hat{u}_{1}^{sc},\cdots,  \hat{u}_{m}^{sc})$.
		\item Generate level one errors within each area $j$ by independently drawing simple random samples of size $N_j$ with replacement from $\hat{\boldsymbol{e}}^{sc}_{h(j)}$ where $h(j)$ is a random drawn from $1,\dots,m$. Denote this sample by $\boldsymbol{e}^{\ast}_{j}$.
	\end{enumerate}
	\item Calculate the vector of pseudo-random effects $\hat{g}_j=\bar{\boldsymbol{x}}_j(\hat{\boldsymbol{\alpha}}_{\hat{\bar{q}}_j}- \hat{\boldsymbol{\alpha}}_{0.5})$ from the M-quantile model for binary data, where $\bar{\boldsymbol{x}}_j$ is the vector of average values for the covariates in area $j$. Then center around zero and re-scale the empirical pseudo-random effects. Define the new vector as $\hat{\boldsymbol{g}}^{cs}=(\hat{g}_1^{cs},\dots,\hat{g}_m^{cs})$.
	\item Generate a vector of random intercepts $\hat{\boldsymbol{g}}^{\ast}$ for $m$ areas by independently drawing simple random samples of size $m$ with replacement from the vector $\hat{\boldsymbol{g}}^{cs}$.
	\item Simulate bootstrap population data $(y_{ij}^{\ast},\mathbf{x}_{ij}, w^{\ast}_{ij})$:
	$$y_{ij}^{\ast}=\mathbf{x}_{ij}^{T}\hat{\boldsymbol{\beta}}_{0.5}+w^{\ast}_{ij}\hat{\gamma}_{0.5} +w^{\ast}_{ij}\gamma_{j}^{\ast}+u_{j}^{\ast}+\epsilon_{ij}^{\ast},$$
	where $ w_{ij}^{\ast} \sim \operatorname{Bernoulli}(p_{ij}^{\ast})$ and $ p_{ij}^{\ast}= \Lambda^{-1}(\mathbf{x}_{ij}^{T}\hat{\boldsymbol{\alpha}}_{0.5} + g^{\ast}_{j})$.
	\item Draw a bootstrap sample from the bootstrap population using the sampling method used to obtain the original sample.
	
	\item Fit the models in the equations \eqref{P-eq:estMQ1}, \eqref{P-eq:estMQ2} and predict for the non-sampled part of the bootstrap population. Then calculate the bootstrap estimated outcomes $\hat{y}_{ij}^{\ast}$ and propensity score $\hat{p}_{ij}^{\ast}$. 
\item Calculate the bootstrap errors $\hat{\tau}_{j}^{\ast}-\tau_{j}^{\ast}$, for each $j=1, \cdots, m$.  The value $\tau_{j}^{\ast}$ is the true bootstrap value obtained by replacing $\hat{y}_{ij}^{\ast}$ and $p_{ij}^{\ast}$ in equation \eqref{eq:PATE3} and the $\hat{\tau}_{j}^{\ast}$ is the bootstrap estimate and it is derived by replacing the $\hat{y}_{ij}^{\ast}$ and $\hat{p}_{ij}^{\ast}$ in the same equations. 
	\item Repeat steps $4-9$ $B$ times. In the $b$th bootstrap replication, let $\tau_{j}^{\ast(b)}$ be the quantity of interest for area $j$ and $\hat{\tau}_{j}^{\ast(b)}$ its bootstrap estimate. A bootstrap estimator of area $j$ for the variability contribution of the estimates of propensity scores to the MSE of IPW-MQ can be calculated as
	\begin{equation}
	    \var(\hat{\tau}_{j})=\frac{1}{B}\sum_{b=1}^{B}(\hat{\tau}_{j}^{\ast(b)}-\tau_{j}^{\ast(b)})^2.
	\end{equation}
\end{enumerate} 
 This value has to be added to the estimator of the MSE in equation \eqref{P-MQ_predictor_final}. 
 
While fitting a line on the marginal residuals it might happen that $\hat{\gamma}_{j}$ as in (\ref{gamma_boot}) is infinite. This is the case if for instance there are no treated units in the sample. In such cases, the classical methods that will only use the sample for inferences are incapable of providing an estimate of average treatment for such areas. However, our proposal can still provide an estimate in such cases by using the information on the entire sample. To solve this problem in our proposed bootstrap procedure in the area $j$-th where there is no treated unit in the sample, we restrict the $\hat{\gamma}_{j}=0$ and then automatically $\hat{u}_{j}=\hat{r}_{.j}$. This makes sense since the sample data are not informative about the area specific effect of the treatment.

\section{Benchmarking properties of the estimators} \label{sec:prop}
Model-based methods may not satisfy coherence properties, that may be relevant to final
users of small area estimates. In this section we focus on the benchmarking property of the proposed small area predictors. Let the small
areas be a partition of a larger area. A set of estimates is said to be benchmarking if the
estimated average of the treatment effect for the small areas sum to the average estimated for the larger area (typically using design unbiased or design consistent methods).

The IPW-EBLUP and IPW-MQ do not satisfy the benchmarking property. The estimator of the average treatment effect of the entire population is $\hat{\tau}_{PATE}$, and this may be written as the weighted average of the small area predictors $\hat{\tau}_{PATE_{j}}$. This property is desirable for all small area estimators, and it is, in this case, an alternative statement of the benchmarking property. In this section the weights that guarantee the benchmarked properties of the proposed estimators are obtained. 

At population level the $\hat{\tau}_{PATE}$ can be written as follow: 
\begin{align}\label{eq:total}
    \hat{\tau}_{PATE}= \sum_{j=1}^{m}\sum_{i \in s_{j}}\mathscr{D}_{ij}y_{ij} +\sum_{j=1}^{m}\sum_{i \in r_{j}}\mathscr{D}_{ij}\hat{y}_{ij},
\end{align}
where $ \mathscr{D}_{ij}= \left(\frac{\mathscr{K}^{-1}w_{ij}}{\hat{e}(\mathbf{x}_{ij})}-\frac{\mathscr{T}^{-1}(1-w_{ij})}{1-\hat{e}(\mathbf{x}_{ij})}\right),$
 $\mathscr{K}= \sum_{j=1}^{m}\sum_{i=1}^{N_{j}}w_{ij}/\hat{e}(\mathbf{x}_{ij})= \sum_{j=1}^{m}K_{j}$ and   $\mathscr{T}=\sum_{j=1}^{m}\sum_{i=1}^{N_{j}}(1-w_{ij})/(1-\hat{e}(\mathbf{x}_{ij}))=\sum_{j=1}^{m}T_{j}$.
The benchmarking equation can be written as $\hat{\tau}_{PATE}=\sum_{j=1}^{m}\mathscr{A}_{j}\hat{\tau}_{PATE_{j}} $, where $\mathscr{A}_{j}$s are the weights indicating the contribution of the average treatment effect of area $j$ to the average treatment effect of the entire population. Therefore, it follows that:
  \begin{align*}
     \hat{\tau}_{PATE} & =\sum_{j=1}^{m}\mathscr{A}_{j}\hat{\tau}_{PATE_{j}}\\
                       & = \sum_{j=1}^{m}\mathscr{A}_{j}\left( \sum_{i\in s_{j}}D_{ij}y_{ij}+\sum_{i\in r_{j}}D_{ij}\hat{y}_{ij}\right)\\
                       & = \sum_{j=1}^{m} \sum_{i\in s_{j}}\mathscr{A}_{j} D_{ij}y_{ij}+\sum_{j=1}^{m} \sum_{i\in r_{j}}\mathscr{A}_{j}D_{ij}\hat{y}_{ij}.
 \end{align*}

 Then the weight for area $j$ is 
 
 \begin{equation}
    \mathscr{A}_{j}= \frac{\mathscr{D}_{ij}}{D_{ij}} \qquad  \forall i \in \mathcal{U}_{j},
 \end{equation}
 
 where $\mathscr{A}_{j}$ is a unique solution $ \iff \mathscr{A}_{j}=K_{i}/\mathscr{K}=T_{j}/\mathscr{T}$. This means that to guarantee the benchmarking property the weighted percentage of treated and controlled outcomes(weighted by their propensities) in each area to the total population of treated and controlled outcomes, respectively, must be equal to the proportion of the sub-population size in that area to the total population size. 
 
 An alternative solution to satisfy the benchmarking property for the proposed estimators is to consider weighing the two parts of equation (\ref{P-eq:PATE3}) differently. In this case we define the average treatment effect of the entire population as :
 \begin{align*}
     \hat{\tau}_{PATE} & =\sum_{j=1}^{m} \mathscr{B}_{j}\left. \left(\sum_{i\in s_{j}}\left[\frac{w_{ij}y_{ij}}{\hat{e}(\mathbf{x}_{ij})}\right]+\sum_{i\in r_{j}}\left[\frac{w_{ij}\hat{y}_{ij}}{\hat{e}(\mathbf{x}_{ij})}\right]\right) \left(\sum_{i=1}^{N_{j}}\frac{w_{ij}}{\hat{e}(\mathbf{x}_{ij})} \right)^{-1} \right. - \nonumber \\
&
\sum_{j=1}^{m}\mathscr{C}_{j}\left. \left(\sum_{i\in s_{j}}\left[\frac{(1-w_{ij})y_{ij}}{1-\hat{e}(\mathbf{x}_{ij})}\right]+\sum_{i\in r_{j}}\left[\frac{(1-w_{ij})\hat{y}_{ij}}{1-\hat{e}(\mathbf{x}_{ij})}\right]\right)\left(\sum_{i=1}^{N_{j}}\frac{1-w_{ij}}{1-\hat{e}(\mathbf{x}_{ij})}\right)^{-1}\right. ,
\end{align*}
where $\mathscr{B}_{j}=K_{i}/\mathscr{K}$ and $\mathscr{C}_{j}=T_{j}/\mathscr{T}$.

\section{Model-based simulations }\label{sec:supp_Simulation}
The validity of model-based inference depends on the validity of the model assumed. In
this section we empirically evaluate the properties of small area predictors and corresponding
MSE estimators. In particular, we use Monte Carlo simulation to evaluate the performance of the proposed small area estimators in comparison with the performance of the IPW-Direct estimator at small area level. Our simulations are model-based, in the
sense that population data are first generated under a model assumption or scenario, with a
sample selected from each simulated population. Estimates of small area effects and
corresponding MSEs are computed by using the data from these samples. Population data are generated for $m=50$ small areas, with samples selected
by simple random sampling without replacement within each area. The population and sample
sizes are the same for all areas and are fixed at either $N_i =100$ and $n_i =5$ or $N_i =300$ and
$n_i =15$. The auxiliary information, the values of the covariates and the treatment status are known for all the units in the population. The response variable ($y_{ij}$) and the propensity score (the probability of being treated, $e(\mathbf{x}_{ij})$) have been generated as
\begin{equation}\label{eq:simy}
y_{ij}= 100+2x_{1ij}+x_{2ij}+\tau_{j}w_{ij}+u_{j}+\epsilon_{ij},
\end{equation}
\begin{equation}\label{eq:sime}
e(\mathbf{x}_{ij})=\frac{exp(-1+0.5 x_{2ij}+\nu_{j})}{1+exp(-1+0.5 x_{2ij}+\nu_{j})},
\end{equation}
where $x_{1ij} \sim \operatorname{LogNormal}(1,\,0.5)$, $x_{2ij} \sim \operatorname{Uniform}(0,\, 1)$,  $\mathbf{x}_{ij}=(x_{1ij},x_{2ij})$, and $w_{ij} \sim \operatorname{Bernoulli}(e(\mathbf{x}_{ij})).$ The random-area and individual effects
are independently generated according to four scenarios:
\begin{enumerate}
    \item no outliers, no misclassification: $u_{j} \sim \mathcal{N}(0, \, 3)$, $\epsilon_{ij} \sim \mathcal{N}(0, \, 6)$ and $\nu_{j}\sim \mathcal{N}(0, \, 0.25)$ is the baseline. The other scenarios are modified upon this setting; 
    \item outliers in both area and individual effects: $u_{j} \sim \mathcal{N}(0, 3)$ for areas $1-39$ and $u_{j}\sim \mathcal{N}(9,20)$ for areas $40-50$, that is, random effects for areas $1-39$ are drawn from a ‘well-behaved’ $\mathcal{N}(0, 3)$ distribution, with those for areas $40-50$ drawn from an outlier $\mathcal{N}(9, 20)$ distribution; $\epsilon_{ij} \sim \delta_{1}\mathcal{N}(0,  6)+(1-\delta_{1}) \mathcal{N}(20,150)$ where $\delta_{1}$ is an independently generated Bernoulli random variable with $Pr(\delta_{1})=0.97$, that is, the individual effects are independent draws from a mixture of two normal distributions, with $97\%$ on average drawn from a ‘well-behaved’ $\mathcal{N}(0, 6)$ distribution and $3\%$ on average drawn from an outlier $\mathcal{N}(20, 150)$ distribution;
    \item misclassification in treatment status: $w_{ij}=w_{ij}'\delta_{2}+(1-w_{ij}')(1-\delta_{2})$, where $w_{ij}' \sim \operatorname{Bernoulli}(e(\mathbf{x}_{ij}))$ and $\delta_{2}$ is an independently generated Bernoulli random variable with $Pr(\delta_{2})=0.98$;
    \item outliers in both area and individual effects and misclassification in treatment status: $u_{j} \sim \mathcal{N}(0, \, 3)$ for areas $1-39$ and $u_{j}\sim \mathcal{N}(9, \, 20)$ for areas $40-50$, $\epsilon_{ij} \sim \delta_{1} \mathcal{N}(0, \, 6)+(1-\delta_{1})\mathcal{N}(20, \, 150)$ and $w_{ij}=w_{ij}'\delta_{2}+(1-w_{ij}')(1-\delta_{2})$. 
\end{enumerate}
    Each scenario includes two sub-scenarios that introduce a different degree of heterogeneity in the effects among areas: a) $\tau_{j} \sim \mathcal{N}(10, \, 1)$ and b) $\tau_{j} \sim \mathcal{N}(10,3)$. Each scenario is independently simulated $S=1000$ times. Table \ref{tab:Scenarios} summarises the scenarios of the simulation experiments.

\begin{table}[H]
	\begin{center}
		\caption{Model based simulation settings}
		\label{tab:Scenarios}
		\begin{tabular}{|l|c|c|c|c|}
			\hline
			 Scenarios & $\tau_{j}\sim \mathcal{N}(10,1)$ & $\tau_{j}\sim \mathcal{N}(10,3)$ & Presence of outliers & Misclassification\\\hline
			 $1-a$ &  \checkmark & & &  \\\hline
			 $1-b$ &  &\checkmark & & \\\hline
			 $2-a$ & \checkmark & &\checkmark &  \\\hline
			 $2-b$ &  &\checkmark &\checkmark&  \\\hline
			 $3-a$  &\checkmark & &  &\checkmark  \\\hline
			 $3-b$  & &\checkmark & &\checkmark \\\hline
			 $4-a$   &\checkmark & & \checkmark &\checkmark  \\\hline
			 $4-b$  & &\checkmark & \checkmark  &\checkmark \\\hline
		\end{tabular}
	\end{center}
\end{table}

Three different estimators are compared: IPW-Direct (\ref{P-eq:SATE3}); IPW-EBLUP (\ref{P-eq:PATE3}) using equations (\ref{P-eq:est2}) and (\ref{P-eq:est3}) to predict the outcomes and to estimate the propensity scores, respectively; IPW-MQ (\ref{P-eq:PATE3}) with equations (\ref{P-eq:estMQ1}) and (\ref{P-eq:estMQ2}) for computing outcomes and propensity scores, respectively.

\begin{figure}[H]
\centering
\includegraphics[width=1\linewidth]{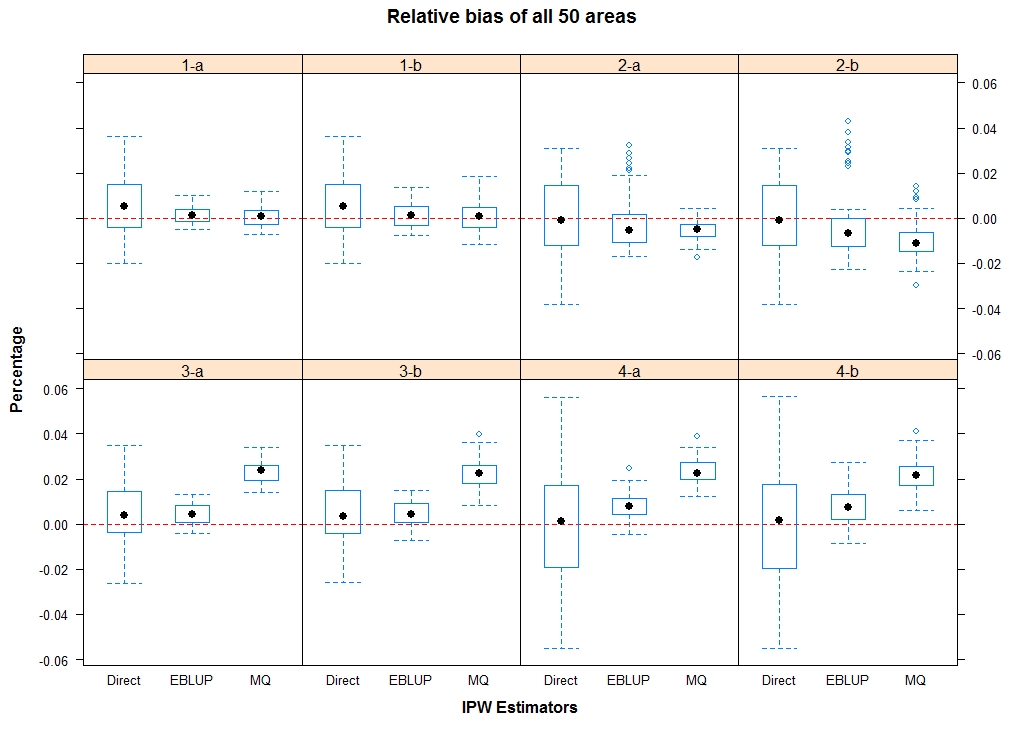}
\caption{Distribution of the median values of area-specific relative bias (computed over simulations) in percentage for different estimators in the setting with $N_{j}=100$ and $n_{j}=5$. Each boxplot represents the percentage of relative bias in estimating the average treatment effect. Note that the Direct, EBLUP and MQ stand for IPW-Direct, IPW-EBLUP and IPW-MQ, respectively.}
\label{fig:RB-100}
\end{figure}

\begin{figure}[H]
\centering
\includegraphics[width=1\linewidth]{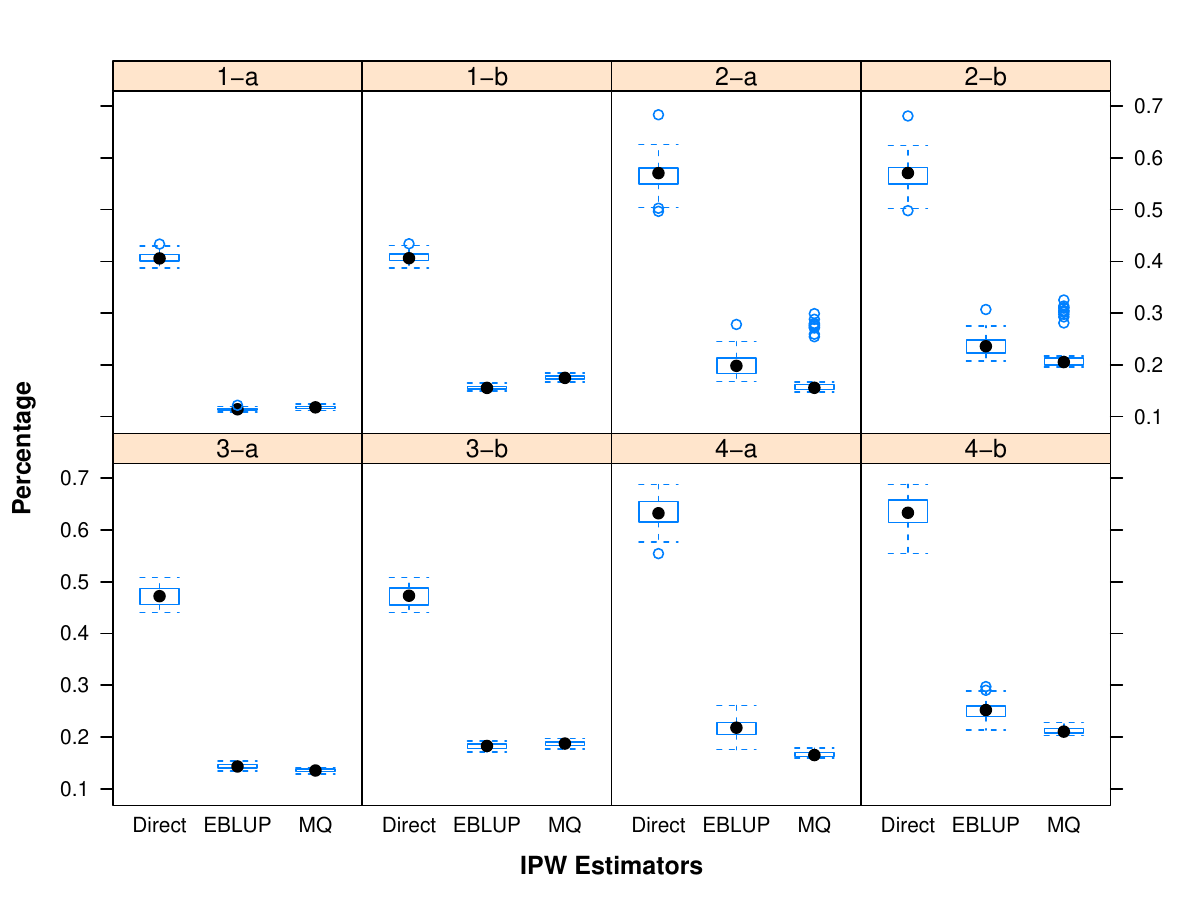}
\caption{Distribution of the median values of area-specific relative root mean square errors (computed over simulations) in percentage for different estimators in the setting with $N_{j}=100$ and $n_{j}=5$. Each boxplot represents the percentage of RRMSE in estimating the average treatment effect. Also note that the Direct, EBLUP and MQ stand for IPW-Direct, IPW-EBLUP and IPW-MQ, respectively.}
\label{fig:RRMSE-100}
\end{figure}

The performances of these different small area estimators for area $j$ were evaluated with
respect to two criteria: the percentage of Relative Bias (RB) and the percentage of Relative Root Mean Square Errors (RRMSE):
 $$ RB_{j}= \bar{\tau}_{j}^{-1}\frac{1}{S}\sum_{s=1}^{S}(\hat{\tau}_j^{s}-\tau_{j}^{s})\times 100,$$
 
 $$ RRMSE_{j}=\bar{\tau}_{j}^{-1} \sqrt{\frac{1}{S}\sum_{s=1}^{S}\left(\hat{\tau}_{j}^{s}-\tau_{j}^{s}\right)^2} \times 100, $$
 where $\bar{\tau}_{j}=\frac{1}{S}\sum_{s=1}^{S}\tau_{j}^{s}$, $\hat{\tau}_{j}^{s}$ is the estimate of the effect $\tau_j^s$ in area $j$ for iteration $s$. The distribution of the median values (across simulations) of area-specific RB and RRMSE are set out by boxplots in Figures \ref{fig:RB-100} and \ref{fig:RRMSE-100}, where we see that claims in the literature \citep{chambers2006} about the superior outlier robustness of the M-quantile predictor compared with the IPW-Direct and the IPW-EBLUP certainly hold true in these simulations.
 
 The relative bias results confirm our expectations regarding the behaviour of the estimators: the IPW-Direct is less biased than the model-based predictors IPW-EBLUP and IPW-MQ. In particular, the increase in bias is most pronounced in IPW-MQ when there are misclassification and outliers in the area and individual effects, which is not unexpected since IPW-MQ is a robust estimator and area means are most affected by outliers in the population data \citep{Chambers2014}. To reduce the bias a predictive estimator could be developed adding a bias correction part in line with that proposed by \citet{Chambers2014}. A predictive IPW-MQ is an avenue for future research.
 
 Turning to the median RRMSE results, the superior outlier robustness of IPW-MQ compared with the IPW-EBLUP certainly hold true — provided that the outliers are in individual and in area effects (scenarios $2$ and $4$). However, the gap between these two estimators narrows considerably when only misclassification is present (scenarios $3-a$ and $3-b$). Nevertheless, the proposed small area predictors, IPW-EBLUP and IPW-MQ, are much more efficient than the IPW-Direct and this suggests that it may be good to use these predictors to estimate the average treatment effect when sample size in each area becomes small. These results are confirmed when $N_j=300$ and $n_j=15$ with lower values of RB and RRMSE for all estimators. For reasons of space, they are not reported here, but are available from the authors upon request.

%
%

%
%
%

To evaluate the performance of the analytic MSE estimators of the IPW-EBLUP and IPW-MQ proposed in Section \ref{P-sec:MSE} (see equations \eqref{P-eq:mse-eblup} and \eqref{P-MQ_predictor_final}) we used the data generated for scenarios $1$ and $4$ with sample size $n_j =5$. Scenario $1$ is shown because it represents the situation under which the MSE estimator for IPW-EBLUP has been developed. Instead scenario $4$ has been chosen to assess the robust properties of the MSE estimators of both predictors. The results for the other scenarios are available for interested readers from the authors upon request. The performance of the MSE estimators has been evaluated by median values of Relative Bias of root MSE over the $m$ small areas and the median of the empirical coverage rate (CR) for nominal $95\%$ confidence intervals. {\color{black} They are defined by the small area estimates plus or minus twice the values of the estimated root  MSE}. Examination of
the results in Table \ref{tab:MSE-RB} shows that the MSE estimators \eqref{P-MQ_predictor_final} tend to be biased towards low values, except for the case $4-a$. This confirms the results obtained by \citet{Chambers2014}. The MSE estimators \eqref{P-eq:mse-eblup} show good performance in terms of bias under all scenarios. In terms of CR the two MSE estimators show poor results with rates around $85-90\%$. However, this use of the
estimated MSE to construct confidence intervals, though widespread, has been criticised. \citet{Chatterjee:2008} discuss the
use of bootstrap methods for constructing confidence intervals for small area parameters, arguing that there is no guarantee that the asymptotic behaviour underpinning normal theory confidence intervals applies in the context of the small samples that characterise small area estimation. Further research on using bootstrap and/or jackknife techniques to construct confidence intervals under the mixed effects and M-quantile models is left for the future. 

\begin{table}[H]
    \begin{center}
	\caption{Median values (across the areas) of the Relative Bias and Coverage Rate ($\%$) (computed over simulations) for the root MSE estimators of IPW-EBLUP and IPW-MQ under scenarios $1$ and $4$ with $N_{j}=100$, $n_{j}=5$.}
	\vspace{0.5cm}
	\label{tab:MSE-RB}
\begin{tabular}{lrrrr}
\hline
\multirow{2}{*}{Scenario} & \multicolumn{2}{c}{IPW-EBLUP}      &  \multicolumn{2}{c}{IPW-MQ}\\ \cline{2-5}
 & {RB\%}      & CR (95\%) & {RB\% }  & CR (95\%) \\\hline
1-a       & -0.83 & 88   & -4.84 & 89           \\
1-b      & -4.73 & 91      & -33.64 & 75      \\
4-a       & -1.38 & 89         & 3.88 & 93     \\
4-b       & -8.59 & 86      & -16.47 & 86       \\\hline
\end{tabular}
\end{center}
\end{table}

\section{Some analysis of the real data used for the  design-based simulation experiment} \label{sec:supp_app}
\subsection{Data}
\begin{figure}[ht]
	\centering
	\includegraphics[width=0.8\linewidth]{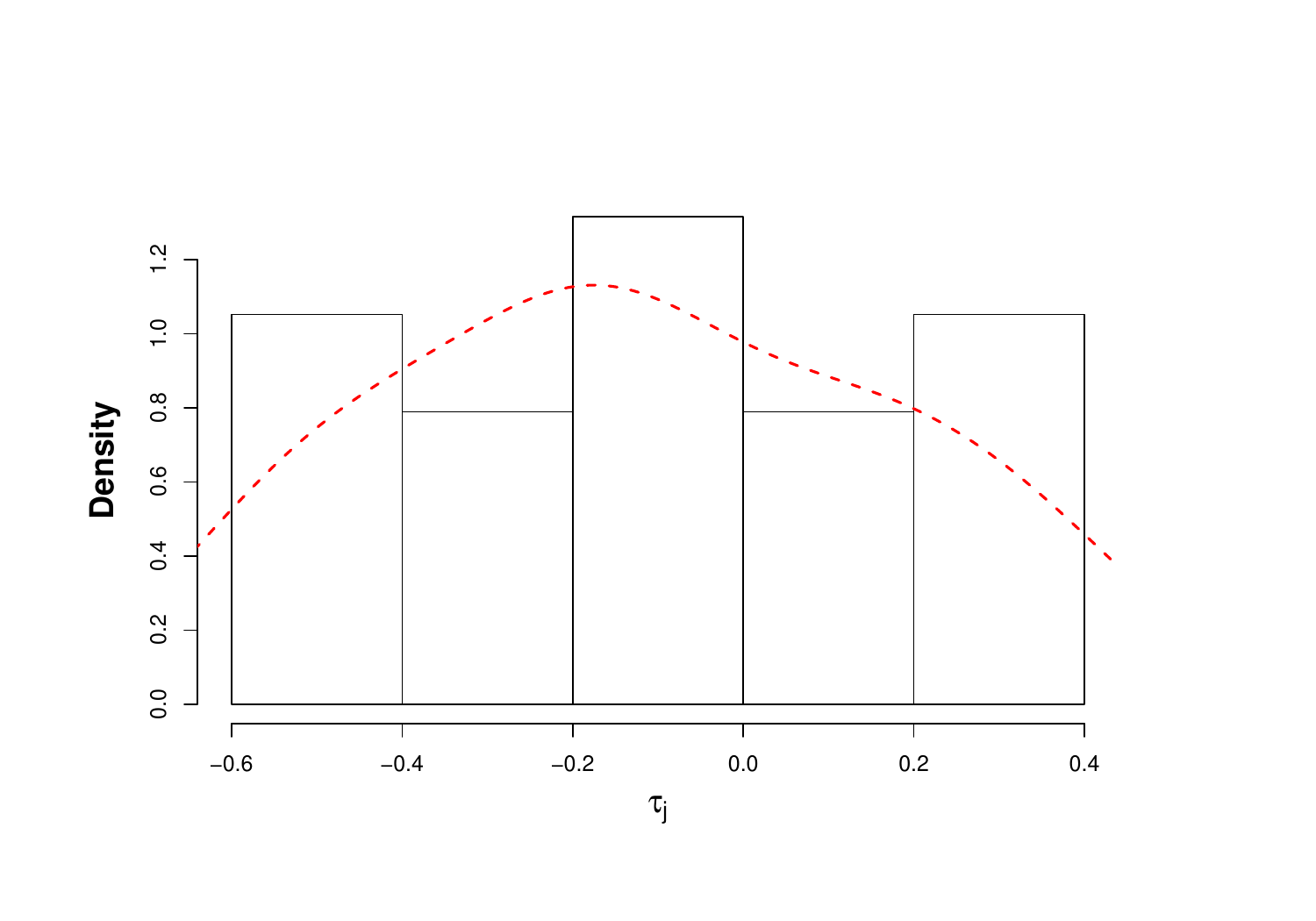}
	\caption{Distribution of area-specific population average treatment effect, showing the heterogeneity of the effects.}
	\label{fig:hist_true}
\end{figure}

Table \ref{tab:Orig_sample} shows the full sample parameters. The heterogeneity of the area-specific effects is illustrated in Figure \ref{fig:hist_true}. 

\begin{table}[H]
	\centering
	\caption{Parameters of the full sample, to be considered as population parameters.}
	\label{tab:Orig_sample}
	\begin{tabular}{lrrr}
		\hline
		NUTS2                 & Original sample size & Percentage of treated & Original estimate (True) \\\hline
		Piemonte & 665 & 0.07 & -0.37 \\ 
		Valle-d'Aosta & 152 & 0.13 & -0.43 \\ 
		Lombardia & 1329 & 0.04 & -0.13 \\ 
		Bolzano-Trento & 444 & 0.14 & -0.48 \\ 
		Veneto & 927 & 0.08 & -0.16 \\ 
		Friuli-Venezia-Giulia & 735 & 0.10 & -0.31 \\ 
		Liguria & 598 & 0.08 & 0.37 \\ 
		Emilia-Romagna & 842 & 0.13 & -0.24 \\ 
		Toscana & 707 & 0.08 & -0.55 \\ 
		Umbria & 310 & 0.12 & -0.59 \\ 
		Marche & 586 & 0.14 & -0.16 \\ 
		Lazio & 934 & 0.12 & 0.10 \\ 
		Abruzzo-Molise & 363 & 0.10 & 0.21 \\ 
		Campania & 594 & 0.29 & -0.13 \\ 
		Puglia & 481 & 0.25 & 0.37 \\ 
		Basilicata & 200 & 0.31 & 0.11 \\ 
		Calabria & 398 & 0.44 & 0.23 \\ 
		Sicilia & 475 & 0.18 & -0.13 \\ 
		Sardegna & 271 & 0.19 & 0.06 \\ \hline
		Italy & 11011 & 0.13 & -0.14 \\ 
		\hline
	\end{tabular}
\end{table}
\subsection{Common support and balance in the covariates}
We assume the availability of the auxiliary data on the treatment status and the confounding covariates via other sources of information. This allows us to check the assumption of common support at the sub-population level for each area and correct for the unmatched items. In order to reach the complete overlap of the propensity scores between the treated and control unit within each area, we dismissed 1349 in the overall pseudo-population, ending up with 11011 units for the analysis. 
\begin{figure}[H]
\centering
\includegraphics[width=0.6\linewidth, height=8cm]{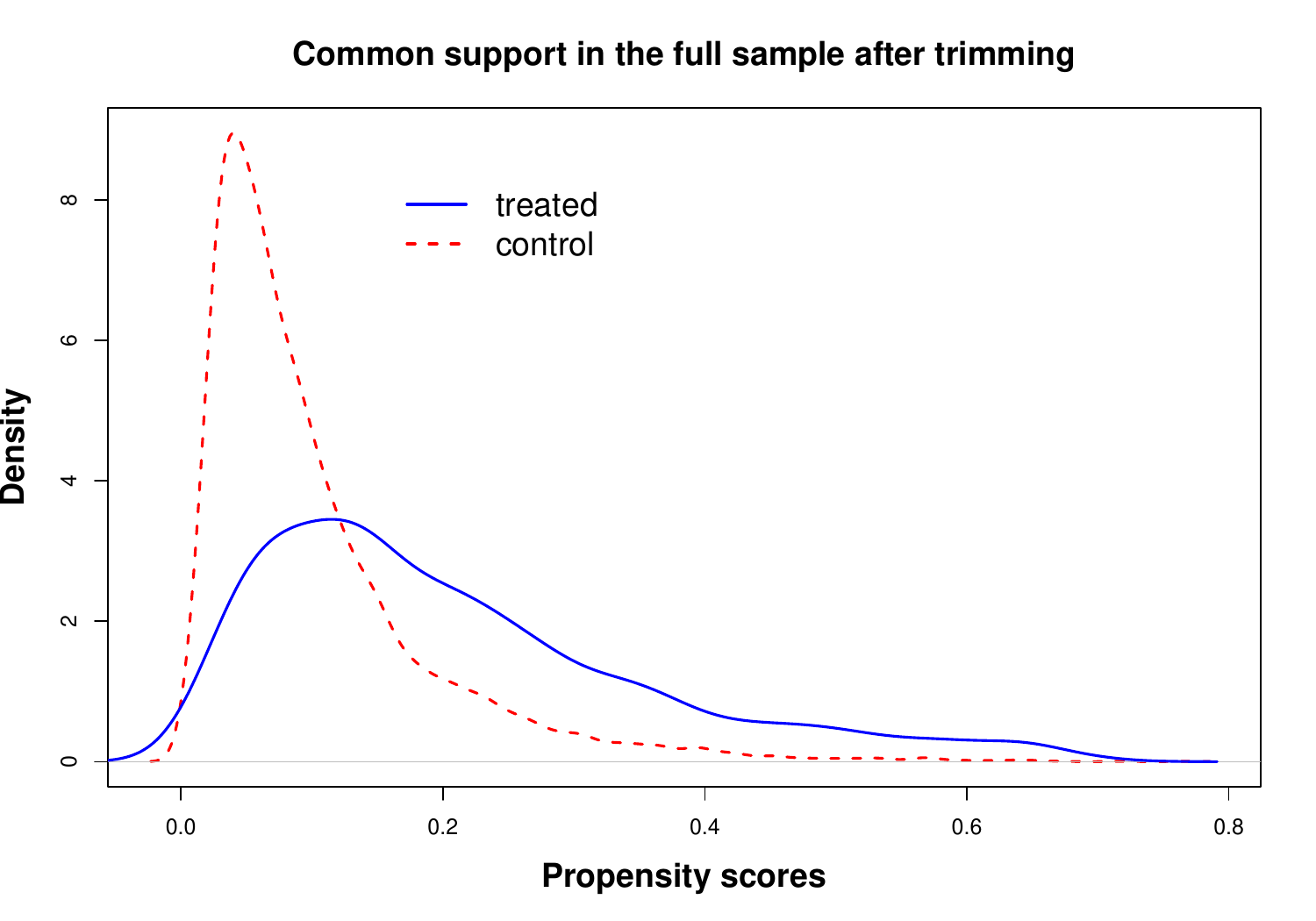}
\caption{The overall common support based on propensity scores.}
\label{fig:com-supl}
\end{figure}

We check the balancing property in the set of covariates using the propensity scores for the entire pseudo-population.
The NULL hypothesis ($H_0$) is that the distribution of the covariates is balanced in the control and treated group within area $j$.

Consider, $l(\mathbf{x}_{ij})$, the linearized propensity score or the odds of being treated, is defined as follows:
$$l(\mathbf{x}_{ij})=\ln \left(\frac{e(\mathbf{x}_{ij})}{1-e(\mathbf{x}_{ij})}\right)$$
where $e(\mathbf{x}_{ij})$ is the true propensity score. Now we define the $\bar{l}_{cj}$ and the $\bar{l}_{tj}$ as the average value of linearized propensity score for the control and treated group in area $j$, respectively. 

The following test statistic can be used to verify if weighting with these scores provides the balance required between the treated and control in each region. It follows then a Student's $t$ distribution:

$$ \hat{\Delta}_{j}=\frac{\bar{l}_{t}-\bar{l}_{c}}{\sqrt{(s^{2}_{l_{c}}+s^{2}_{l_{t}})/2}} \sim t_{\nu}.$$

The number of degrees of freedom, $\nu$, is calculated as:
$$ \nu = \frac{(s^{2}_{l_{c}}/N_{j}^{c}+s^{2}_{l_{t}}/N_{j}^{t})^2}{(s^{2}_{l_{c}}/N_{j}^{c})^2/(N_{j}^{c}-1)+(s^{2}_{l_{t}}/N_{j}^{t})^2/(N_{j}^{t}-1)},$$
where $N_{j}^{c}$ and $N_{j}^{t}$ are the sizes of treated and control sub-populations in area $j$. 

Table \ref{tab:balance} reports the results of this test within each area, showing that  $H_{0}$ cannot be rejected for any area $j$, $j=1, \cdots, m$.

\begin{table}[H]
\centering
\caption{Testing the balance of covariates within each area using the linearized propensity scores}\label{tab:balance}
\vspace{2mm}
\begin{tabular}{lcc}
  \hline
NUTS2 & Test statistic &  p-value \\ 
  \hline
Piemonte & 0.77 & 0.44 \\ 
  Valle-d'Aosta & 0.58 & 0.57 \\ 
  Lombardia & 0.73  & 0.47 \\ 
  Bolzano-Trento & 0.56 & 0.58 \\ 
  Veneto & 0.82 & 0.41 \\ 
  Friuli-Venezia-Giulia & 0.84 & 0.40 \\ 
  Liguria & 0.97 & 0.34 \\ 
  Emilia-Romagna & 0.63  & 0.53 \\ 
  Toscana & 0.56 & 0.57 \\ 
  Umbria & 0.60 &  0.55 \\ 
  Marche & 0.81 &  0.42 \\ 
  Lazio & 0.80 &  0.43 \\ 
  Abruzzo-Molise & 0.49 & 0.63 \\ 
  Campania & 0.92 & 0.36 \\ 
  Puglia & 0.91 &  0.36 \\ 
  Basilicata & 0.48  & 0.63 \\ 
  Calabria & 0.88 &  0.38 \\ 
  Sicilia & 0.55 &  0.58 \\ 
  Sardegna & 0.91 & 0.36 \\ 
   \hline
\end{tabular}
\end{table}

\bibliographystyle{chicago}
\bibliography{CISAE}